%% file: main.tex
\documentclass{IEEEoj}
\input{setup/preamble}

\begin{document}

\input{sec/00-frontmatter}

\maketitle

\input{sec/01-introduction}

\input{sec/02-motivation}

\input{sec/03-platoon-model}
\input{sec/04-controller-design}

\input{sec/05-numerical-evaluation}

\input{sec/06-experimental-validation}
\input{sec/07-conclusion}
\input{sec/08-outlook}

\input{sec/90-appendix}

\bibliographystyle{IEEEtran}
\bibliography{refs}

\end{document}

%% file: setup/preamble.tex
\usepackage{cite}                      %
\usepackage{amsmath,amssymb,amsfonts}
\usepackage{algorithmic}
\usepackage{graphicx,color}
\usepackage{textcomp}
\usepackage{xcolor}
\usepackage{array}                     %
\usepackage{booktabs}                  %
\usepackage{url}
\usepackage{stfloats}                 %
\usepackage{multirow}                  %
\usepackage{makecell}

\def\BibTeX{{\rm B\kern-.05em{\sc i\kern-.025em b}\kern-.08em
    T\kern-.1667em\lower.7ex\hbox{E}\kern-.125emX}}
\AtBeginDocument{\definecolor{ojcolor}{cmyk}{0.93,0.59,0.15,0.02}}
\def\OJlogo{}



\makeatletter
\def\@maketitle{\newpage
  \bgroup%
    \vspace*{-10.5pt}%
     \vskip14.5pt{\titlefont{\textcolor{ojcolor}{\@title}}\par\vskip8.5pt}%
    {\authorfont\@author\par}\vskip4pt{\afffont\@affil}\par\@IEEEspecialpapernotice\vskip5pt%
    \ifx\@corresp\@empty\else{\corfont\@corresp\par}\vskip5.5pt\fi%
    \ifx\@authornote\@empty\else{\afffont\@authornote\par}\vskip27.5pt\fi%
    \centerline{\rule{36pc}{.5pt}}\par\vskip11.5pt%
    \ifvoid\abstractbox\else{\reset@font\box\abstractbox}\fi\par\vskip11.5pt%
    \ifvoid\keybox\else{\reset@font\box\keybox}\fi\par\vskip-32pt%
    \centerline{\rule{36pc}{.5pt}}%
    \par\addvspace{33.5pt}\egroup%
    \let\maketitle\relax
}%
\def\ps@plain{%
   \def\@oddhead{\vbox{\hsize\textwidth\vspace*{-13pt}\vbox to 29pt{\hsize126pt\vspace*{11.5pt}\OJlogo}\hfill\par\vspace*{-2.75pt}\hbox to \textwidth{\vrule width\textwidth height.3pt depth0pt}}}%
   \let\@evenhead\@oddhead%
   \def\@evenfoot{\hbox to \textwidth{{\rffont\thepage}\hfill}}%
   \def\@oddfoot{\hbox to \textwidth{\hfill{\rffont\thepage}}}%
}%
\def\ps@headings{%
   \def\@oddhead{\vbox{\hsize\textwidth\vspace*{-13.5pt}\hfill\hbox{\vbox to 29pt{\hsize122.8pt\vspace*{12pt}\OJlogo}}\par\vspace*{-2.25pt}\hbox to \textwidth{\vrule width\textwidth height.3pt depth0pt}}}%
   \def\@evenhead{\vbox{\hsize\textwidth\vbox to 0pt{\hsize\textwidth\vspace*{7.7pt}\rfxfont\raggedright\rightmark:\ \leftmark\hfill\par}\par\vspace*{16pt}\hbox to \textwidth{\vrule width\textwidth height.3pt depth0pt}}}%
   \def\@evenfoot{\hbox to \textwidth{{\rffont\thepage}\hfill}}%
   \def\@oddfoot{\hbox to \textwidth{\hfill{\rffont\thepage}}}%
}%
\makeatother

%% file: sec/00-frontmatter.tex
\title{Cooperative Adaptive Cruise Control with Constant Distance Gaps
for Retrofitted Legacy Vehicles}

\author{KAY MASSOW\affilmark{1},
EMANUELE CRISOSTOMI\affilmark{2},
ILJA RADUSCH\affilmark{3}, AND
ROBERT SHORTEN\affilmark{4}}
\affil{Daimler Center for Automotive IT Innovations, Technische
Universit\"at Berlin, 10587 Berlin, Germany}
\affil{Department of Energy, Systems, Territory and Constructions
Engineering, University of Pisa, Pisa, Italy}
\affil{Fraunhofer Institute FOKUS, Berlin, Germany}
\affil{Dyson School of Design Engineering, Imperial College London,
SW7 2AZ London, U.K.}
\corresp{CORRESPONDING AUTHOR: Kay Massow (e-mail: kay.massow@dcaiti.com).}
\authornote{Preprint. A condensed version of this work has been submitted to the IEEE Open Journal of Intelligent Transportation Systems.}

\markboth{Cooperative Adaptive Cruise Control with Constant Distance Gaps
for Retrofitted Legacy Vehicles}{Massow \textit{et al.}}

\begin{abstract}
A Constant Distance Gap (CDG) spacing controller for Cooperative Adaptive Cruise Control (CACC) designed to increase the capacity of oversaturated signalized intersections is presented. In a departing platoon of production passenger vehicles, the controller maintains close inter-vehicle distances below 3~m within limited platoon lengths, using one-vehicle-look-ahead (OVLA) communication only. Extensive real-world experiments on heterogeneous, retrofitted legacy production cars demonstrate robust operation. To the best of our knowledge, this is the first experimental demonstration of such close distances under this combination of conditions. A CDG policy can substantially increase discharge rates at oversaturated signalized intersections when vehicles preserve their standstill spacing during startup, as shown in our previous simulation study. Because this benefit unfolds only at high penetration rates, it calls for retrofit-capable solutions deployable in existing combustion-dominated fleets. However, practical realization in such fleets remains an open challenge. The proposed MPC-based controller combines a first-order longitudinal prediction model with precise powertrain feedforward employing OEM-specific torque models, without exchanging confidential vehicle parameters. It avoids explicit platoon coordination to minimize air-interface load in dense traffic. To mitigate disturbance amplification and increase feasible platoon length at very small gaps, two controlled spacing-relaxation strategies are compared: (i) a micro-time-gap policy near the theoretical string-stability bound, and (ii) an acceleration-dependent spacing relaxation based on a virtual mass--spring--damper model. We introduce the Desired Acceleration Overshoot (DAO), a time-domain metric relating local actuator-demand amplification to achievable platoon length.
\end{abstract}


%% file: sec/01-introduction.tex
\section{Introduction}\label{sec:intro}

\textsc{C}ooperative Adaptive Cruise Control (CACC) can increase traffic efficiency and road capacity \cite{shladover_2015_definitions,ploeg_2014_thesis}, particularly at oversaturated and traffic-light-controlled intersections \cite{massow_2022_numerical,vannunen_2012_competition}. By integrating V2V communication with onboard sensors, CACC enables vehicles to operate at shorter inter-vehicle distances than conventional ACC \cite{ploeg_2011_design}.

A central design choice is the spacing policy. While most CACC implementations adopt the Constant Time Gap (CTG) policy because of its favorable string-stability properties \cite{ploeg_2014_thesis}, the resulting inter-vehicle distance increases with velocity and therefore limits traffic density. Constant Distance Gap (CDG) policies can instead provide substantial capacity benefits \cite{swaroop_1999_constant,darbha_2020_multiple,lioris_2017_doublethroughput,timmerman_2021_platoonforming}, especially at low speeds and within limited platoon lengths \cite{massow_2022_numerical}. The capacity gains at signalized intersections depend on safety gaps, response times \cite{yan_2026_startuploss}, and acceleration and braking limits \cite{levine_2016_queuedischarge}.

Our previous work \cite{massow_2022_numerical} showed in vehicle and traffic simulation that, at oversaturated traffic lights, compact CDG platoons can significantly increase intersection discharge rates, particularly when the penetration of equipped vehicles is high. This necessitates the rapid retrofitting of existing heterogeneous fleets, primarily internal combustion (ICE) and hybrid (HEV) vehicles. Furthermore, urban deployment at high penetration rates requires minimizing communication overhead to prevent link congestion. This is addressed via a One-Vehicle-Look-Ahead (OVLA) topology in which each vehicle receives information only from its immediate predecessor, which facilitates retrofit implementation using standard V2X equipment.

However, transitioning from macroscopic traffic simulations to real-world deployment reveals a gap. The controller evaluated in \cite{massow_2022_numerical} relied on high-precision acceleration trajectories derived from vehicle-specific maps. Although originally designed for real-world vehicles, this approach failed on production vehicles with legacy drivetrains, showing spacing errors up to 2~m in extreme situations and poor robustness against unmodeled disturbances and drivetrain latencies.

Translating the potential of CDG shown in simulation \cite{massow_2022_numerical} into practice requires a robust controller that operates under real-world conditions in legacy vehicles, using standard sensors and V2X communication. Improving control quality (variance, min/max error) is required for safe and comfortable operation at gaps below 3~m. Ideally, this is achieved by exchanging existing ACC units, facilitating the market penetration necessary to reap the full potential of CDG. However, the heterogeneity of legacy fleets in drivetrain components (ICE, HEV, EV) and actuator delays demands high controller robustness. In the literature, the smallest gaps are often achieved by delay compensation using Smith predictors \cite{dehaan_2023_actuatordelay,xing_2019_smith} or advanced MPC \cite{vannunen_2019_stringstable,kianfar_2015_controlmatching,gratzer_2022_stringstable,he_2026_markovian}. These approaches often require either distributed control, complex communication structures, or detailed driveline models as part of the controller \cite{navas_2021_multimodel,flores_2023_enabling,milanes_2014_realtraffic,lee_2025_openroad}.

Experimentally demonstrated CACC systems have predominantly employed CTG spacing \cite{ploeg_2011_design,vannunen_2019_stringstable,milanes_2014_realtraffic}, whereas CDG experiments have mainly concerned leader-coordinated highway truck platoons, motivated by aerodynamic drag \cite{davila_2011_sartre,tsugawa_2011_trucksplatoon}, or a passenger-car launch demonstration with larger gaps (approximately 6~m) and leader-broadcast communication \cite{smith_2020_throughput}. Field operational tests on signalized corridors have likewise employed CTG spacing (0.6~s) \cite{calvert_2020_fot}. Whether the constant distance gap below 3~m shown in simulation can instead be held during motion in heterogeneous legacy vehicles has, to the best of our knowledge, not yet been experimentally demonstrated.

This work addresses the gap between the simulation-based potential of CDG-based CACC with such small gaps and its real-world implementation. Our core hypothesis is that for scenarios at oversaturated traffic lights, short platoons with very small spacing should be prioritized when the following circumstances occur at the same time:

\begin{itemize}
\item
  Traffic throughput is the primary objective.
\item
  Platoon sizes are small enough that string instability or communication topology complexity can be handled, e.g. by employing mini-platoon control \cite{massow_2022_numerical}, for up to 20 vehicles as motivated in \cite{massow_2024_closerange}.
\item
  Velocities are low enough to meet safety requirements, while maintaining smooth and predictable acceleration.
\end{itemize}

We demonstrate that inter-vehicle distances below 3~m can be maintained in heterogeneous production vehicles under these conditions. By achieving a transient spacing error of less than 20~cm during startup in experiments, we enable the formation of compact mini-platoons \cite{swaroop_1999_constant}, significantly increasing the discharge rate at signalized intersections. The main contribution is maintaining such very short inter-vehicle distances in heterogeneous production vehicles using predecessor-only communication, by reusing the production sensors and actuators together with the manufacturer's torque interface over the vehicle-internal CAN. No hardware beyond that of a CTG-based CACC implementation is required. Before being retrofitted into the test vehicles, the controller was developed and tuned in the PHABMACS simulator \cite{massow_2018_prototyping}, calibrated to the target vehicles.

The OVLA topology is chosen, i.e., each vehicle receives information only from its immediate predecessor. This allows for redundancy through visual or radar-based line-of-sight verification and remains robust against packet loss by minimizing air-interface load, further simplifying the retrofitting of stock production vehicles. We deliberately accept the limited string stability of CDG \cite{darbha_2020_multiple} and treat the resulting amplification of disturbances as a design constraint on the admissible platoon length. The main contributions of this work are:

\begin{enumerate}
\item
  Development and evaluation of a robust MPC-based CDG controller: The controller uses a simple first-order prediction model together with precise OEM-level powertrain feedforward and is specifically designed for retrofitting heterogeneous legacy production vehicles.
\item
  Systematic comparison of relaxation strategies: We evaluate two controlled spacing-relaxation strategies (a Micro Time Gap policy and an acceleration-dependent spacing relaxation) to minimize disturbance amplification along the platoon, quantifying maximum feasible platoon sizes in simulation.
\item
  Experimental validation: The system is implemented and evaluated on production passenger vehicles with different powertrain characteristics (ICE, HEV), providing quantitative performance metrics and deriving illustrative estimates of platoon-size limits.
\item
  Mini-Platoon Strategy \cite{swaroop_1999_constant} for traffic-light-controlled intersections: We intentionally prioritize small-gap efficiency over strict string stability and larger time gaps as design goals \cite{ploeg_2014_thesis,flores_2023_enabling,vannunen_2017_robust}, using the OVLA topology as a compromise between performance, real-world feasibility, and retrofitting effort.
\end{enumerate}

The demonstrated practical feasibility bridges the gap between the traffic-level benefits predicted for compact CDG platoons and their practical implementation on existing vehicles.

The remainder of this paper is structured as follows: Section~\ref{sec:relatedwork} reviews the state of the art.
Sections~\ref{sec:platoon-model} and~\ref{sec:controller} detail the platoon model and MPC design. Sections~\ref{sec:numeval} and~\ref{sec:exp} present the numerical analysis and the experimental validation, followed by the conclusion in Section~\ref{sec:conclusion}.

%% file: sec/02-motivation.tex
\section{Related Work and Preliminary Considerations}\label{sec:relatedwork}

This section motivates the one-vehicle-look-ahead communication pattern for CDG spacing and reviews related work on heterogeneous drivelines, actuator dynamics, and delay compensation in CACC platoons. It then derives the two spacing-relaxation strategies investigated in this work: the Micro Time Gap policy and the acceleration-dependent spacing relaxation.

\subsection{Motivation of One-Vehicle-Look-Ahead Communication Pattern for CDG Spacing }\label{sec:motivation}

Before proceeding, some clarifying remarks on the use of the one-vehicle-look-ahead communication pattern for CDG and its implications for string stability are in order. This pattern is often referred to as the predecessor--follower (PF) topology in the literature \cite{li_2015_overview}. Beyond strict string stability, weaker notions have been introduced that permit local or limited amplification while requiring the overall response to remain bounded \cite{ploeg_2014_lpstring}. Earlier work has shown that CDG can achieve string stability when each vehicle receives information from the platoon leader or from multiple predecessors \cite{darbha_2020_multiple}. He et al.\ \cite{he_2026_markovian} address string stability by switching between different communication topologies. In this study, however, we deliberately accept the potential limitations of PF-based string stability for the following reasons.

\begin{enumerate}
\item
  PF does not require a formal platoon architecture, i.e., designating a leader and explicitly ordering vehicles in a platoon \cite{sheikholeslam_1993_longitudinal}. Such an architecture demands reliable mutual identification of non-adjacent vehicles without direct line-of-sight and a standardized set of platooning protocols. Achieving this is considerably more difficult than maintaining simple unidirectional communication with the direct predecessor.
\item
  For CDG, reliable low-latency communication is required. Receiving real-time information from a platoon leader with multiple sending vehicles in the middle implies a higher rate of packet drops than from the direct predecessor. Moreover, relying on simultaneous real-time information from multiple vehicles inherently increases the error rate compared to processing data from a single upstream source \cite{vannunen_2019_stringstable}. Ma et al.\ \cite{ma_2025_timeheadway} quantify the effect of such errors.
\item
  Identification in PF can be supported visually, e.g. by encoding vehicle IDs in the rear lights of vehicles, which can be detected by the follower's front camera. Real-time data transmission could also be augmented through Visible Light Communication (VLC) \cite{plattner_2023_visiblelight} and Radar-based Communication technologies \cite{schettler_2019_visiblelight}.
\end{enumerate}

To bound instabilities in the platoon, we restrict operation to finite mini-platoons \cite{swaroop_1999_constant} and treat the admissible platoon length as a design variable, while relying solely on predecessor information. This limitation allows for an implementation without explicit platoon coordination by propagating the number of preceding vehicles upstream \cite{massow_2024_closerange}. Specifically, each vehicle includes in its transmitted data the count received from its predecessor, incremented by one. Once this propagated count exceeds the desired maximum platoon length, that vehicle ``splits'' the platoon by switching from CDG to a CTG spacing policy.

\subsection{Heterogeneous drive lines, actuator dynamics, and delay compensation in CACC-Platoons}\label{sec:heterogeneous-drivelines}

The main challenge in realizing very small spacing in CACC platoons is the heterogeneity of powertrain dynamics across different vehicle types. This heterogeneity arises not only from different vehicle masses and dimensions, but to a large extent from different powertrain topologies \cite{guanetti_2018_control} and components (ICE, HEV, EV, gearboxes, torque converter states, actuator delays) \cite{ersal_2020_connected}. In the literature, therefore, accurate powertrain models are considered for numerous cooperative functions, such as eco-driving, coordinated control of engine actuators with traffic light preview, fuel consumption optimization, and ride comfort \cite{ersal_2020_connected,guanetti_2018_control}. Detailed powertrain models and precise low-level controllers are employed to compensate not only for heterogeneity, but also for measurement noise and disturbances such as road grade changes, wind, or rolling resistance \cite{viadero_2024_robust}.

One line of work explicitly addresses delays in platoons without requiring knowledge of the preceding vehicle's driveline model. In \cite{dehaan_2023_actuatordelay}, actuator delay is compensated using a Smith predictor (in simulation). In \cite{dehaan_2025_heterogeneous}, an integrated controller directly accounts for delays in the controller design and is validated in real-vehicle experiments. However, smaller inter-vehicle gaps can be achieved when the driveline model of the preceding vehicle is explicitly exploited. Examples include multi-model adaptive control for heterogeneous platoons \cite{navas_2021_multimodel} and approaches based on nonlinear vehicle models with bidirectional asymmetric control algorithms \cite{monteil_2019_linfinity}. In \cite{wang_2015_stringstable}, guidelines are provided for selecting the inter-vehicle spacing as a function of driveline heterogeneity within the platoon. Huisman et al.\ \cite{huisman_2026_plantequivalent} address parameter uncertainty on the controller side instead, tuning alternative realizations of the same nominal controller rather than relying on detailed plant knowledge.

While many of these approaches are validated predominantly in simulation, real-world experiments with homogeneous platoons show that relatively simple PD-like controllers can achieve time gaps of 0.7~s \cite{ploeg_2011_design}, and, with delay compensation, of about 0.25~s \cite{xing_2019_smith}. MPC-based approaches achieve experimentally validated time gaps of 0.3~s in heterogeneous platoons \cite{vannunen_2019_stringstable} (to the best of our knowledge, the smallest time gap demonstrated with real heterogeneous vehicles; see their Table III).

Drivelines of legacy ICE vehicles are particularly difficult to model and are used in real-world experiments in \cite{milanes_2014_realtraffic}, \cite{kapsalis_2025_cutins}, and \cite{flores_2023_enabling}, typically based on a second-order model commonly employed for ICE driveline modeling. Liu et al.\ \cite{liu_2023_lowerlevel} focus explicitly on the application of such a second-order model for precise low-level control in real-vehicle tests. The work of Flores et al.\ \cite{flores_2023_enabling} is one of the few that aim at precise control of a real-world CACC platoon with different powertrain types (hybrid electric and ICE) and is therefore particularly relevant for retrofitting CACC to legacy vehicles. The approach employs actuator maps and a second-order low-level controller. Kapsalis et al.\ \cite{kapsalis_2025_cutins} adopt the hierarchical CACC framework of \cite{flores_2023_enabling} and extends the platoon by including EV powertrains.

For delay compensation in CACC common approaches are the Smith-predictor \cite{dehaan_2023_actuatordelay}, \cite{xing_2019_smith} or Model Predictive Control (MPC) \cite{vannunen_2019_stringstable}, \cite{kianfar_2015_controlmatching}. Communication delays impose different requirements than actuator delay on the ego vehicle, because the follower only observes delayed predecessor acceleration and therefore never has access to its instantaneous value. In \cite{xing_2020_compensation}, the feedforward communication delay is compensated by introducing a master--slave CACC architecture that places the delay in series with the controlled plant, thereby enabling the application of a Smith predictor. De Haan et al.\ \cite{dehaan_2023_actuatordelay} compensate actuator delay on the ego vehicle using a Smith predictor and adjust the time gap to correct the resulting error in steady-state situations. Wijnbergen et al.\ \cite{wijnbergen_2026_decentralized} embed actuator and communication delays directly in a delayed spacing policy without using a Smith predictor, rather than treating them as disturbances to be compensated. MPC-based CACC \cite{vannunen_2019_stringstable}, \cite{kianfar_2015_controlmatching} handles delays predictively by embedding actuator and communication delays directly into the prediction model and optimizing future inputs over the horizon. Robustness to packet loss can be further improved with prediction buffers \cite{vannunen_2019_stringstable}.

In summary, small inter-vehicle gaps in heterogeneous platoons were achieved by combining (i) controller-side delay compensation \cite{xing_2020_compensation}, \cite{vannunen_2019_stringstable} with (ii) precise drivetrain control \cite{flores_2023_enabling}, \cite{kapsalis_2025_cutins}. For ICE vehicles, second-order low-level models are well suited \cite{liu_2023_lowerlevel}, but harder to integrate into a unified formal framework than first-order models. In this work, we therefore adopt an MPC-based spacing controller using a first-order vehicle model for prediction, while precise drivetrain control is realized by feedforward exploiting detailed driveline knowledge -- an approach similar to \cite{vannunen_2019_stringstable} and \cite{kianfar_2015_controlmatching}. The idea is not to increase controller complexity, as done, e.g., in \cite{zheng_2017_dmpc}, \cite{huisman_2026_plantequivalent}, but to shift complexity toward the vehicle-level drivetrain control, which is already available in production vehicles. This separation allows the spacing controller to remain structurally simple while maintaining high tracking precision in heterogeneous powertrains. Moreover, this approach works without sharing (confidential) knowledge on vehicle characteristics between the vehicles.

\subsection{Derivation of a Micro Time Gap Policy for Comparison with CDG}\label{sec:microgap-derivation}

As motivated earlier, we employ CDG to maintain the standstill distance between vehicles when starting at traffic lights. As one possible method of damping spacing errors, we introduce a Micro Time Gap policy as a specific relaxation method, which follows a CTG law but uses a very small, nonzero time gap. The core idea behind this relaxation is to provide just enough temporal flexibility to reduce the accumulation of spacing errors without significantly increasing the inter-vehicle distance and, thus, to increase the feasible length of the platoon. To define this threshold, we deliberately select a time gap that aligns with the formal lower bounds for string stability under the predecessor--follower (PF) communication pattern, thereby pushing the strategy to its theoretical limit. For that purpose, a literature review on the fundamental limits of string-stable time gaps was conducted.

Many studies have formally derived the minimum string-stable time gap as a function of the information flow topology in the platoon. In general, this gap depends on the communication delay \(\theta\), the lag \(\tau\), and the actuation delay \(\phi\) of the vehicle model. For the PF (OVLA) case considered in this work, Darbha et al.\ \cite{darbha_2019_benefits}, for example, showed that using acceleration, velocity and position for a controller with an acceleration feedback gain \(k_{a} = 1\) leads to a lower bound for the time gap of \(h_{\min} = \tau_{o}\). Here, \(\tau_{o}\) is the maximum actuation lag of a platoon while desired accelerations and communication delays were neglected. Abolfazli et al.\ \cite{abolfazli_2022_timeheadway} included the communication delay in the work of Bian et al.\ \cite{bian_2019_reducing}, \cite{bian_2018_multiplepredecessor}, inspired by \cite{darbha_2019_benefits}, which results for the PF case in \(h_{\min} = 2\left( \tau_{i} + \theta \right)/(2k_{a} + 1)\). Ma et al.\ \cite{ma_2025_robust} treat \(\tau\) as an uncertain parameter with \(h_{\min} = \tau + \theta\) if \(\tau > \theta\).

While most works on finding the lower bounds of time gaps include a controller structure in the analysis, Besselink et al.\ \cite{besselink_2017_delaybased} address string stability as the result of the spacing policy rather than the specific controller design. Besselink et al.\ \cite{besselink_2017_delaybased} propose a delay-based spacing policy (DBSP) with \(s_{ref,i}(t) = s_{i - 1}(t - 1\Delta t)\) where vehicle \(i\) tracks a time-delayed version of the trajectory of vehicle \(i-1\), with the delay \(\Delta t\), which results in equal velocity profiles \(v_{i}(s) = v_{i - 1}(s)\) in space \(s\). Wijnbergen et al.\ \cite{wijnbergen_2023_decentralized} extended this concept to lateral control and showed that the CTG policies can be interpreted for practical implementation as a first-order Taylor approximation of the DBSP. Wijnbergen et al.\ \cite{wijnbergen_2026_decentralized} systematically integrated actuation delays into DBSP and proposed the ``delayed extended spacing policy'' with the referenced spacing \(\Delta_{i}^{ref}(t) = h_{v}v_{i}(t) + h_{a}a_{i}\left( t + \phi_{i} \right)\). \(h_{v}\) denotes the velocity-dependent time gap and \(h_{a}\) acts as an acceleration-dependent term designed to compensate for the vehicle's actuator lag \(\phi_{i}\) directly within the spacing reference.

\begin{figure}[!t]
  \centering
  \includegraphics[width=0.95\columnwidth]{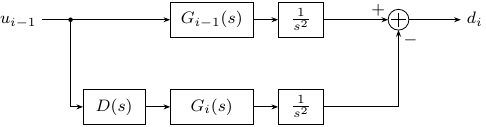}
  \caption{Block diagram of the Micro Time Gap policy in the frequency domain. Vehicle \(i\) tracks the a time-delayed acceleration trajectory of vehicle \(i-1\).}
  \label{fig:micro-gap-block}
\end{figure}

Inspired by Besselink et al.\ \cite{besselink_2017_delaybased}, we use a simplified thought experiment to motivate the potentially smallest string-stable time gap as a delay-based spacing policy. In contrast, we consider PF (OVLA) architectures only, where the following vehicle \(i\) tracks a time-delayed version of the acceleration trajectory of vehicle \(i - 1\). Additionally, we employ prediction in the policy by including \(u_{i - 1}\). For this purpose, we propose the following thought experiment, depicted as a block diagram in Fig.~\ref{fig:micro-gap-block}.

When vehicle \(i - 1\) starts moving from standstill with an input signal \(u_{i - 1}\), let the earliest point in time when vehicle \(i\) notices that movement be delayed by the communication delay \(\theta\). The smallest string-stable gap is obtained when vehicle \(i\) exactly mimics the acceleration of vehicle \(i - 1\). This is achieved for an input signal \(u_{i}(t) = u_{i - 1}(t - \theta)\) assuming \(\tau_{i - 1} = \tau_{i}\) and \(\phi_{i - 1} = \phi_{i}\) for simplification. The models \(G_{i}\) and \(G_{i - 1}\) represent the dynamics of the ego vehicle model and vehicle \(i - 1\) in the frequency domain, where \(s\) is the Laplace operator. \(D\) represents the communication delay \(\theta\) and the input \(u_{i - 1}\) of \(G_{i - 1}\) can be regarded as the desired acceleration. If we assume a constant acceleration during the time span \(\theta\), the resulting inter-vehicle distance \(d_{i}\) (standstill distance neglected) in the time domain reads:
\begin{equation}\label{eq:micro-time-gap}
d_{i}(t) = v_{i}(t) \cdot \theta + a_{i}(t) \cdot \tfrac{1}{2}\theta^{2}
\end{equation}
or
\begin{equation}\label{eq:micro-time-gap-pred}
d_{i}(t) = v_{i - 1} \cdot \theta - a_{i - 1} \cdot \tfrac{1}{2}\theta^{2}
\end{equation}
Assuming a maximum acceleration of \(10\,\mathrm{m/s^2}\), the overall upper bound is \(v_{i} \cdot \theta + 5\theta^{2}\), which could be reduced to \(v_{i - 1} \cdot \theta\) for acceleration and \(v_{i} \cdot \theta\) for braking.

To the best of our knowledge, the smallest string-stable time gap validated in simulation was 0.09~s \cite{xing_2020_compensation} using a bidirectional communication pattern. The smallest string-stable time gap validated experimentally was 0.25~s \cite{xing_2019_smith} using a PF (OVLA) communication pattern. In both cases, a communication delay of 0.04~s, a communication update frequency of 25~Hz, and homogeneous vehicle parameters were applied.

\subsection{Derivation of an Acceleration-Dependent Spacing Relaxation}\label{sec:asr-derivation}

The primary factor inducing instabilities in constant spacing regimes is the delayed reaction of vehicle \(i\) to the acceleration changes of vehicle \(i - 1\), predominantly caused by the communication delay. To compensate for the resulting distance error, vehicle \(i\) is typically forced to overshoot the acceleration of its predecessor. While this overshooting cannot be fundamentally eliminated in rigid systems, it can be mitigated by strategically permitting a transient spacing error.

\begin{figure}[!t]
  \centering
  \includegraphics[width=\columnwidth]{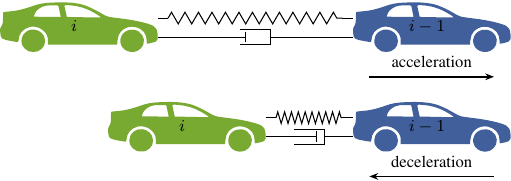}
  \caption{Mechanical analogy of the spacing relaxation modeled as a virtual mass-spring-damper system.}
  \label{fig:msd-analogy}
\end{figure}

One method to achieve this is through soft tuning of the spacing controller. However, this approach reduces tracking precision and increases vulnerability to sensor noise. A more precise alternative is to integrate a dynamic relaxation directly into the spacing policy. This ensures that the controller can maintain high gains for the tracking error -- ensuring zero steady-state error -- while permitting controlled deviations during transients.

\begin{figure}[!b]
  \centering
  \includegraphics[width=0.62\columnwidth]{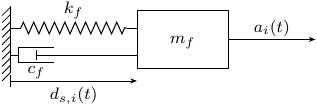}
  \caption{The spacing relaxation component \(d_{s,i}(t)\) represents the response to the driving input acceleration \(a_{i}(t)\), governed by the design parameters \(m_f\), \(k_{f}\) and \(c_f\)}
  \label{fig:msd-response}
\end{figure}

Our approach implements this relaxation by modeling the dynamic component of the inter-vehicle distance as a virtual mass--spring--damper system, as depicted in Fig.~\ref{fig:msd-analogy}. In this model, the vehicle's own acceleration \(a_{i}(t)\) acts as the driving input, causing the desired gap \(d_{s,i}(t)\) to expand elastically during acceleration phases (see Fig.~\ref{fig:msd-response}). This reduces the acceleration overshoot required to enforce rigid spacing during the transient.

In contrast to \cite{wijnbergen_2026_decentralized}, where acceleration enters the desired spacing through a static term, the proposed formulation shapes the spacing reference through a second-order transient response. It should therefore be interpreted as dynamic reference shaping rather than as a model of nonlinear spacing evolution or a physical interaction force.

The system is described by the following second-order differential equation:
\begin{equation}\label{eq:asr-transfer}
\frac{d_{s,i}(s)}{a_{i}(s)} = \frac{K_{a}}{m_fs^{2} + c_fs + k_{f}}
\end{equation}
The corresponding state-space representation is given by:
\begin{equation}\label{eq:spacing-relaxation}
\begin{bmatrix}
{\dot{d}}_{s,i}(t) \\
{\ddot{d}}_{s,i}(t)
\end{bmatrix} = \begin{bmatrix}
0 & 1 \\
-\frac{k_{f}}{m_f} & -\frac{c_f}{m_f}
\end{bmatrix}\begin{bmatrix}
d_{s,i}(t) \\
{\dot{d}}_{s,i}(t)
\end{bmatrix} + \begin{bmatrix}
0 \\
\frac{K_{a}}{m_f}
\end{bmatrix}a_{i}(t)
\end{equation}
The subscript \(f\) denotes the virtual spacing-relaxation filter. To ensure a smooth transition without oscillating behavior, the parameters \(m_f\), \(c_f\), and \(k_{f}\) are selected to achieve a damped response, providing a stable and predictable expansion of the desired gap. This formulation enhances robustness by allowing high prioritization of spacing, even in the presence of sensor noise, while permitting smooth, controlled deviations during rapid transients.

Several works employ virtual spring-damper concepts \cite{yanakiev_1996_simplified}, \cite{mirbakhsh_2023_springmass}, \cite{xie_2023_springdamping} as inter-vehicle coupling models to smooth spacing errors and improve platoon stability, in which a controller computes a force-like control input based on spacing and velocity differences. In contrast, our approach does not represent an interaction force. Instead, we shape a desired spacing trajectory through a damped second-order dynamic, directly excited by the ego-vehicle's acceleration, providing a controlled, acceleration-dependent spacing relaxation.

%% file: sec/03-platoon-model.tex
\section{Platoon Model}\label{sec:platoon-model}

\begin{figure}[!t]
  \centering
  \includegraphics[width=\columnwidth]{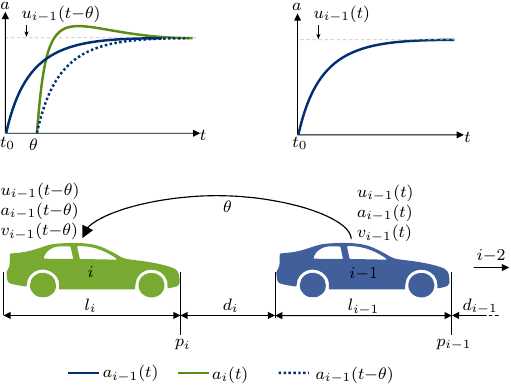}
  \caption{CACC platoon architecture showing transmitted parameters subject to a communication delay \(\theta\). To maintain a Constant Distance Gap (CDG), vehicle \(i\) must compensate for this delay via a transient acceleration overshoot}
  \label{fig:platoon-architecture}
\end{figure}

Consider a platoon of vehicles as depicted in Fig.~\ref{fig:platoon-architecture}, where \(p_{i}(t),v_{i}(t)\), and \(a_{i}(t)\) denote the position (middle of the front bumper), velocity, and acceleration, and \(u_{i}(t)\) the desired acceleration of the \(i_{th}\ \)vehicle in the platoon (hereafter also referred to as ego vehicle) at the time\(\ t\).

Since the goal is to maintain a strictly constant spacing between vehicles at very short distances, each vehicle should ideally match the velocity and acceleration of its predecessor at all times: \(v_{i}(t) = v_{i - 1}(t),\) \(a_{i}(t)\) =\(a_{i - 1}(t)\). In other words, the ego vehicle should replicate the acceleration trajectory of the preceding vehicle. Figure~\ref{fig:platoon-architecture} shows how this is complicated by the communication delay and by the transient acceleration overshoot required to compensate it.

A wireless link is used to share information (\(a_{i - 1}(t),v_{i - 1}(t)\), and \(u_{i - 1}(t)\)) in an OVLA fashion. The transmitted data is subject to a communication delay \(\theta\) caused by queueing, contention, transmission, and propagation \cite{lefeber_2020_heterogeneous} in the network. When vehicle \(i - 1\) initiates an acceleration change at time \(t\), \(t + \theta\) is the earliest time the ego vehicle can start reacting. To compensate for the delayed start and still maintain constant spacing, \(a_{i}(t)\) must temporarily overshoot \(a_{i - 1}(t)\). This overshoot ensures the closing of the initial gap error introduced by the delay, as illustrated by the green trajectory in Fig.~\ref{fig:platoon-architecture}.

With the time constant \(\tau_{i}\) of a first-order LTI system representing the driveline dynamics of vehicle\(\ i\) and its desired acceleration\({\ u}_{i}\), the acceleration dynamics are represented by the following longitudinal model, as commonly adopted in CACC studies \cite{ploeg_2014_thesis,dehaan_2025_heterogeneous,wijnbergen_2026_decentralized}. \({\dot{a}}_{i}(t)\ {= - \ \frac{1}{\tau_{i}}a}_{i}(t) + \frac{1}{\tau_{i}}u_{i}\left( t - \phi_{i} \right),\ {\dot{v}}_{i} = {\ a}_{i},\ {\dot{p}}_{i} = v_{i}\) . This model is obtained from a more detailed, nonlinear model of the longitudinal vehicle motion using feedback linearization \cite{stankovic_2000_decentralized,sheikholeslam_1993_longitudinal} and entails that the vehicle is equipped with a low-level acceleration controller. The model is extended with a constant actuation delay \(\phi_{i}\) \cite{oncu_2014_networkaware,vannunen_2019_stringstable}.

As a basis for the subsequent controller design, we adopt the following widely used error dynamics \cite{ploeg_2014_thesis,huisman_2026_plantequivalent}. The desired spacing \(d_{r,i}(t)\ \)of vehicle \(i\) to vehicle\(\ i - 1\) is given by\(\ d_{r,i}(t) = r_{i} + h_{d,i}v_{i}(t)\), where \(r_{i}\) is the standstill distance, \(l_{i}\) is the length and \(h_{d,i}\) is the desired time gap of vehicle\(\ i\). In the case of constant spacing considered here, we set \(h_{d,i} = 0\). The actual inter-vehicle distance reads \(d_{i}(t) = p_{i - 1}(t) - p_{i}(t) - l_{i - 1}\) and the inter-vehicle distance error\(\ \)is defined as\(\ e_{i}(t) = d_{i}(t) - d_{r,i}(t)\).

We propose to extend these error dynamics in five ways, which form the basis for our subsequent controller design and evaluation:

\begin{enumerate}
\def\labelenumi{\arabic{enumi})}
\item
  \textbf{Redundant information transmission}
\item
  \textbf{Delay compensation via time shifting}
\item
  \textbf{Acceleration-dependent spacing relaxation}
\item
  \textbf{Micro time gap policy}
\item
  \textbf{Jerk-based error dynamics}
\end{enumerate}

\subsection{Redundant information transmission}\label{sec:redundant}

Whereas prior approaches \cite{ploeg_2014_thesis,vannunen_2019_stringstable,oncu_2014_networkaware} typically transmit only \(u_{i - 1}(t)\) (and measure \(d_{i}(t)\) or \({\dot{d}}_{i}(t)\)), we additionally include \(a_{i - 1}(t)\ \)and \(v_{i - 1}(t)\). This redundancy is required for the high tracking precision needed at very short distances, as it mitigates model--plant mismatches in real-world implementations. It enables the controller to directly reference the actual values of these signals, in addition to predicting their future evolution based on model assumptions.

\subsection{Delay compensation via time shifting}\label{sec:delay-comp}

The ego vehicle receives a time-shifted version of the predecessor's acceleration trajectory \(a_{i - 1}(t - \theta)\) (see Fig.~\ref{fig:platoon-architecture}). This temporal shift must be explicitly compensated for, both in the reference values \(a_{i - 1}(t)\ \)and \(v_{i - 1}(t)\) and in the desired acceleration \(u_{i - 1}(t)\), before they can be used as reference values with \(a_{i - 1}(t + \delta)\ \)and \(v_{i - 1}(t + \delta)\). To this end, we introduce a time shift \(\delta\) treated as a configurable parameter for the subsequent controller design.

\subsection{Acceleration-dependent spacing relaxation}\label{sec:spacing-relax}

A controlled trade-off between spacing accuracy and acceleration overshoot (see Fig.~\ref{fig:platoon-architecture}) becomes necessary in platoons with very short distances. To control this trade-off, we introduce a spacing policy that temporarily relaxes the constant spacing constraint by embedding this overshoot directly into the desired distance \(d_{r,i}(t) = r_{i} + h_{d,i}v_{i}(t) + d_{s,i}(t)\) with \(d_{s,i}(t)\) as a function of acceleration, defined in Section~\ref{sec:asr-derivation}.
This formulation enhances robustness by allowing us to maintain strong prioritization of spacing---even in the presence of sensor noise---while still permitting controlled deviations during transients. Note that, throughout the remainder of this work, either \(h_{d,i}\) or \(d_{s,i}\) is set to zero to ensure that only one type of spacing relaxation is active at a time.

\subsection{Micro time gap policy}\label{sec:micro-gap}

For later comparison with a pure constant distance gap (CDG) strategy, we define a Micro Time Gap policy as a relaxation method for CDG (see Section~\ref{sec:microgap-derivation}). It introduces a small, nonzero time gap \(h_{d,i}\) close to the formal lower bound for string stability under the predecessor--follower (PF) communication pattern. This allows us to investigate how minimal temporal spacing, while remaining close to CDG, affects performance and robustness. The delay-based spacing policy of \cite{besselink_2017_delaybased,wijnbergen_2026_decentralized} and the formal lower bounds on string-stable time gaps under predecessor--follower communication \cite{ma_2025_robust} provide the reference points for this choice.

\subsection{Jerk-based error dynamics}\label{sec:jerk}

While prior approaches \cite{ploeg_2014_thesis,vannunen_2019_stringstable,oncu_2014_networkaware} typically focus on the distance error and its first and second derivatives \(e_{i}(t),\ {\dot{e}}_{i}(t),{\ddot{e}}_{i}(t)\), we additionally incorporate the third derivative \({\dddot{e}}_{i}(t)\), i.e., the jerk error. This term damps overshoots in \(u_{i}(t)\) without compromising tracking precision during rapid changes in \(u_{i - 1}(t)\), as may occur with rate-weighted control inputs \cite{vannunen_2019_stringstable}.

\subsection{Vehicle model}\label{sec:vehicle-model}

Building on the five extensions introduced above, we can now refine the underlying vehicle model and define the complete set of error dynamics. In particular, by incorporating the communication delay \(\theta\) and its compensation via the time shift \(\delta\), we obtain the following model for vehicle \(i - 1\):
\begin{equation}\label{eq:vehicle-model}
\begin{split}
{\dot{a}}_{i-1}(t-\theta) = {}& -\frac{1}{\tau_{i-1}}a_{i-1}(t-\theta) \\
 &+ \frac{1}{\tau_{i-1}}u_{i-1}\left( t-\phi_{i-1}-\theta \right)
\end{split}
\end{equation}
Due to the communication delay, the ego vehicle receives a time-shifted version of the predecessor's acceleration trajectory \(a_{i - 1}(t - \theta)\) measured at vehicle \(i\) at time \(t\), which must be explicitly compensated before a and v can be used as reference values with \(a_{i - 1}(t - \theta + \delta) = a_{i - 1}(t)\ \)and \(v_{i - 1}(t - \theta + \delta) = v_{i - 1}(t)\). Assuming that \(\delta\) is small and \({\dot{a}}_{i - 1}(t)\) remains approximately constant over this short interval, we use a first-order Taylor approximation:
\begin{equation}\label{eq:error-dynamics}
a_{i - 1}(t - \theta + \delta) \simeq a_{i - 1}(t - \theta) + \delta{\dot{a}}_{i - 1}(t - \theta)
\end{equation}
This allows us to evaluate the reference signals at time t, and define the error dynamics:
{\small
\begin{equation}\label{eq:state-space}
\begin{bmatrix}
e_{1,i}(t) \\ e_{2,i}(t) \\ e_{3,i}(t) \\ e_{4,i}(t)
\end{bmatrix} =
\begin{bmatrix}
d_{i}(t) - r_{i} - h_{d,i}v_{i}(t) - d_{s,i}(t) \\
v_{i-1}(t-\theta) + \delta a_{i-1}(t-\theta) - v_{i}(t) - {\dot{d}}_{s,i}(t) \\
a_{i-1}(t-\theta) + \delta{\dot{a}}_{i-1}(t-\theta) - a_{i}(t) \\
{\dot{a}}_{i-1}(t-\theta) - {\dot{a}}_{i}(t)
\end{bmatrix}
\end{equation}}
The error states are formulated to optionally account for the predicted predecessor dynamics derived from the communication delay compensation \(\delta\), the dynamic spacing relaxation \(d_{s,i}\), and the time gap \(h_{d,i}\). In this work,\(\ h_{d,i}\) is used exclusively to implement the Micro Time Gap strategy (Sections~\ref{sec:microgap-derivation} and~\ref{sec:micro-gap}). Its application is linked to the simultaneous use of the time shift \(\delta\) (Sections~\ref{sec:microgap-derivation} and~\ref{sec:delay-comp}) on \({\ v}_{i},\ a_{i},\ {\dot{a}}_{i}\). Consequently, \(h_{d,i} = \theta - \delta\) in all configurations employing the time gap policy, while \(h_{d,i} = 0\) otherwise. To allow for independent configuration of these features within the MPC framework, \(h_{d,i}\) is retained as a separate parameter.

The higher-order term of \(d_{s,i}\) in \({\ e}_{3,i}(t)\) is omitted in the controller implementation as it has negligible effect on steady-state tracking while its jerk-driven transient component would unnecessarily disturb the acceleration error signal.

The term \({\ e}_{4,i}(t)\) is treated as a damping term rather than as an error dynamic comparable with \({\ e}_{1,i},\ {\ e}_{2,i},{\ e}_{3,i}\). It includes neither the time shift nor the spacing relaxation.

Although the measured distance is not subject to the communication delay, we add the shift of \(v_{i - 1}\) to the error dynamics, as we need to derive the prediction of the distance error from the velocity error later.\(\ d_{i}(t)\) can alternatively be expressed by:
\begin{equation}\label{eq:output}
d_{i}(t) - r_{i} = \int{v_{i - 1}(t - \theta)} + {\delta v}_{i - 1}(t - \theta) - \int{v_{i}(t)}
\end{equation}

%% file: sec/04-controller-design.tex
\section{Controller Design}\label{sec:controller}

In this section, an MPC controller is designed to minimize the control errors \({\ e}_{1,i}(t),\ {\ e}_{2,i}(t),\ {\ e}_{3,i}(t),\ \)and \(e_{4,i}(t)\). Before presenting the control strategy, we discretize the continuous-time model \eqref{eq:vehicle-model} and the error dynamics. This discretization is performed for accurate controller modeling, since the controller will later operate in discrete time on an embedded system. The discrete vehicle model for vehicle i\(-\)1 is given by:

\[
\begin{aligned}
a_{i-1}\left( k+1-\theta^{d} \right) = {}&
  a_{i-1}\left( k-\theta^{d} \right)\left( 1-\frac{T_{s}}{\tau_{i-1}} \right) \\
 &+ \frac{T_{s}}{\tau_{i-1}}u_{i-1}\left( k-\phi_{i-1}^{d}-\theta^{d} \right)
\end{aligned}
\]
We denote the discretized time by \(k\) and the sampling interval \(T_{s}\), where \(t = kT_{s}\). The actuator delay \(\phi_{i}\) is approximated by the discrete time delay \(\phi_{i}^{d} = \left\lfloor \frac{\ \phi_{i}}{T_{s}} \right\rceil \in N_{0}\), which is the closest integer multiple of \(T_{s}\). Similarly, the discrete representations of \(\phi_{i - 1}\), \(\theta\), and \(\delta\) are defined by \(\phi_{i - 1}^{d}\), \(\theta^{d}\), and \(\delta^{d}\). To ensure dimensional consistency in the discretized error dynamics, we introduce the scaled delay and time-gap parameters \(\overline{\delta} = n_{\delta}\ \theta^{d}T_{s},\ \ n_{\delta} \in Z\) and \({\overline{h}}_{d,i} = n_{h}\ \theta^{d}T_{s},\ \ n_{h} \in N_{0}\).

The state of vehicle \(i\) is described by the state vector:
\[
\begin{aligned}
x_{i}(k) = \big[\,& e_{1,d,i}(k)\ e_{2,i}(k)\ a_{i}(k)\ a_{i-1}\left( k-\theta^{d} \right) \\
 & e_{3,i}(k)\ j_{i}(k)\ e_{4,i}(k)\ \Delta d_{s,i}(k)\ \Delta^{2}d_{s,i}(k) \,\big]^{T}
\end{aligned}
\]

where \(e_{1,i}(k),e_{2,i}(k),e_{3,i}(k),e_{4,i}(k),\) \({\Delta d}_{s,i}(k)\), \(j_{i}(k)\) are the discrete-time expressions of \({\ e}_{1,i}(t),\ {\ e}_{2,i}(t),\ {\ e}_{3,i}(t),\ {\ e}_{4,i}(t)\), \({\dot{d}}_{s,i}(t)\), \({\dot{a}}_{i}(t)\) and \(e_{1,d,i}(k){= e}_{1,i} + {\overline{h}}_{d,i}v_{i}(k)\).\(\ \)

The system dynamics of the open-loop model are obtained by combining the vehicle longitudinal dynamics with the error dynamics:
\[\begin{aligned}
x_{i}(k + 1) = {}& Ax_{i}(k) + B_{1}u_{i}\left( k - \phi_{i}^{d} \right) \\
 &+ B_{2}u_{i - 1}\left( k - \phi_{i - 1}^{d} - \theta^{d} \right)
\end{aligned}\]
\[y_{i}(k + 1) = Cx_{i}(k + 1)\]

\begin{figure*}[!t]
\centering
{\small\setlength{\arraycolsep}{1pt}
\[
A = \begin{bmatrix}
1 & T_{s} & 0 & 0 & \frac{T_{s}^{2}}{2} & 0 & 0 & 0 & - \frac{T_{s}^{2}}{2} \\
0 & 1 & - \left( 1 - \frac{T_{s} - \overline{\delta}}{\tau_{i}} \right)T_{s} & \left( 1 - \frac{T_{s}}{\tau_{i - 1}} \right)T_{s} & 0 & 0 & 0 & 0 & - T_{s} \\
0 & 0 & 1 - \frac{T_{s}}{\tau_{i}} & 0 & 0 & 0 & 0 & 0 & 0 \\
0 & 0 & 0 & 1 - \frac{T_{s}}{\tau_{i - 1}} & 0 & 0 & 0 & 0 & 0 \\
0 & 0 & - \left( 1 - \frac{T_{s} - \overline{\delta}}{\tau_{i}} \right) & 1 - \frac{T_{s}}{\tau_{i - 1}} & 0 & 0 & 0 & 0 & 0 \\
0 & 0 & - \frac{1}{\tau_{i}} & 0 & 0 & 0 & 0 & 0 & 0 \\
0 & 0 & \frac{1}{\tau_{i}} & - \frac{1}{\tau_{i - 1}} & 0 & 0 & 0 & 0 & 0 \\
0 & 0 & 0 & 0 & 0 & 0 & 0 & 1 & T_{s} \\
0 & 0 & 0 & 0 & 0 & \frac{K_{a}T_{s}}{m_f} & 0 & \frac{- T_{s}k_{f}}{m_f} & 1 - \frac{T_{s}c_f}{m_f}
\end{bmatrix},B_{1} = \begin{bmatrix}
0 \\
\frac{- T_{s} + \overline{\delta}}{\tau_{i}}T_{s} \\
\frac{T_{s}}{\tau_{i}} \\
0 \\
\frac{- T_{s} + \overline{\delta}}{\tau_{i}} \\
\frac{1}{\tau_{i}} \\
\frac{- 1}{\tau_{i}} \\
0 \\
0
\end{bmatrix},B_{2} = \begin{bmatrix}
0 \\
\frac{T_{s}}{\tau_{i - 1}}T_{s} \\
0 \\
\frac{T_{s}}{\tau_{i - 1}} \\
\frac{T_{s}}{\tau_{i - 1}} \\
0 \\
\frac{1}{\tau_{i - 1}} \\
0 \\
0
\end{bmatrix},\ C = \begin{bmatrix}
1 & 0 & 0 & 0 & 0 & 0 & 0 & 0 & 0 \\
0 & 1 & 0 & 0 & 0 & 0 & 0 & 0 & 0 \\
0 & 0 & 1 & 0 & 0 & 0 & 0 & 0 & 0 \\
0 & 0 & 0 & 1 & 0 & 0 & 0 & 0 & 0 \\
0 & 0 & 0 & 0 & 1 & 0 & 0 & 0 & 0 \\
0 & 0 & 0 & 0 & 0 & 0 & 1 & 0 & 0
\end{bmatrix}
\]}
\vspace*{4pt}
\hrulefill
\end{figure*}

with the control input \(u_{i}(k)\) being the desired acceleration of vehicle \(i\) and the feedforward input \(u_{i - 1}\left( k - \phi_{i - 1}^{d} - \theta^{d} \right)\) from vehicle \(i - 1\). The direct feedthrough matrix \(D\) is zero and is therefore omitted. The system matrices are defined as \(A,B_{1},B_{2},\ C\).

The output vector \(y_{i}(k) = \ \){[}\({\overline{y}}_{i}(k),e_{4,i}(k)\){]} consists of the measured outputs \({\overline{y}}_{i}(k)\) and the unmeasured output \(e_{4,i}(k)\). Although \(e_{4,i}(k)\) is needed for the MPC cost function, obtaining it from the derivative of the acceleration signal would be too sensitive to noise.

We assume \(T_{s} \geq \theta\), i.e., the sampling interval should be selected slightly above the communication delay. This assumption is required for the CDG case \(\delta^{d} = \theta^{d} > 0\), where stable coefficients are obtained only for \(\overline{\delta} \leq T_{s}\). For the Micro Time Gap case (\(\delta^{d} \leq 0\)), this constraint is inherently satisfied. The communication rate should be higher than \(\frac{1}{T_{s}}\).

\subsection{Prediction Model}\label{sec:prediction}

At every sampling instant \(T_{s}\), the system states and disturbance are measured. Based on these measurements, a Linear Kalman Filter (LKF), described in the next subsection, combines the model prediction with the measurements to obtain the state estimate \({\widehat{x}}_{i}\left( k|k \right)\). The following quadratic programming problem is then solved to determine the optimal control inputs:

\[
\begin{aligned}
\min_{\Delta U_{k}} J(k) = {}&
  \sum_{j=1}^{N_{p}} \left\| \widehat{y}_{i}^{MPC}\!\left( k+j \mid k \right) \right\|_{W_{y}}^{2} \\
\end{aligned}
\]
Here, \(j\) represents future time steps within the prediction horizon \(N_{p}\). The input increments \(\Delta u_{i}(k + j|k) = u_{i}(k + j|k) - u_{i}(k + j - 1|k)\) are stacked into the decision variable \(\Delta U_{k} = \lbrack\Delta u_{i}(k|k),\ldots,\Delta u_{i}(k + N_{c} - 1|k)\rbrack^{T}\) over the control horizon \(N_{c} \leq N_{p}\), beyond which the command is held constant. Only the first increment is applied, i.e., \(u_{i}(k) = u_{i}(k - 1) + \Delta u_{i}(k|k)\), and the optimization is repeated at the next sampling instant.

For the optimization of the control inputs, only the error terms weighted in the cost function are considered, forming the output vector
\[{y_{i}}^{MPC}(k) = \lbrack e_{1,i}(k)\ e_{2,i}(k)\ e_{3,i}(k)e_{4,i}(k)\rbrack^{T}\]
and \({{\widehat{y}}_{i}}^{MPC}\left( k + j \middle| k \right)\) is the predicted value of \({y_{i}}^{MPC}(k)\) at time step \(k + j\). Although \(a_{i}(k)\) and \(a_{i - 1}(k)\) are included in the state vector as a design decision to mitigate model--plant mismatches in the prediction through reference measurements, they receive no weight in the cost function. The corresponding weighting matrix is defined as:
\[W_{y} = \operatorname{diag}\left( w_{e1},w_{e2},w_{e3},w_{e4} \right)\]
\begin{figure*}[!t]
  \centering
  \includegraphics[width=0.82\textwidth]{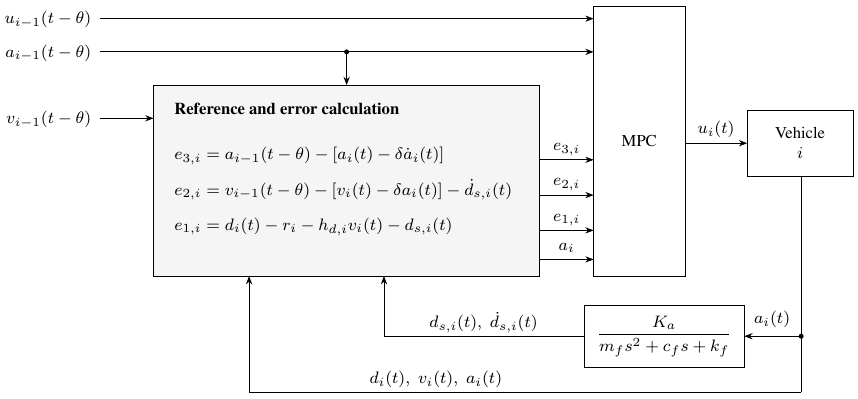}
  \caption{Block diagram of the control loop and the interaction between the MPC and the vehicle plant.
  }
  \label{fig:control-loop}
\end{figure*}

For safety, the predicted spacing and acceleration commands are constrained for all predicted time steps within \(N_{p}\) according to
\begin{equation}\label{eq:mpc-constraints}
\hat e_{1,i}(k+j|k)\geq e_{1,i}^{\min},
\qquad
u_{i,\min}\leq u_i(k+j)\leq u_{i,\max},
\end{equation}

The lower bound \(e_{1,i}^{\min}\) for the predicted spacing error \({\widehat{e}}_{1,i}\left( k + j|k \right)\) ensures a minimum safe distance to the preceding vehicle, and \(u_{i}^{\max}\) is the upper bound for the acceleration command of vehicle \(i\).

\subsection{State Estimation for Practical Implementation}\label{sec:state-est}

To maximize the accuracy of the MPC by incorporating reference measurements from acceleration sensors, a Linear Kalman Filter (LKF) is used to ensure robustness by noise suppression. This enables the practical implementation on a real vehicle, using acceleration sensor data as reference. The LKF estimates the system state \(x_{i}(k)\) at time \(k\), based on measurements \({\overline{y}}_{i}(k)\) available up to the time \(k\) as \({\widehat{x}}_{i}\left( k|k \right)\).

The measurement equation for the LKF reads:
\[{\overline{y}}_{i}(k + 1) = \overline{C}x_{i}(k + 1) + v(k + 1)\]
where\(\ v(k) \sim \ N\ (0,\ R)\) represents the measurement noise and \(\overline{C} \in R^{5x9}\) is the measurement matrix, containing the first five rows from the full system output matrix \(C\) corresponding to the measured outputs. The filter equations are as follows:

Prediction step:

\[
\begin{aligned}
{\widehat{x}}_{i}\left( k+1|k \right) = {}& A{\widehat{x}}_{i}\left( k|k \right) + B_{1}u_{i}\left( k - \phi_{i}^{d} \right) \\
 &+ B_{2}u_{i-1}\left( k-\phi_{i-1}^{d}-\theta^{d} \right)
\end{aligned}
\]
\[P\left( k + 1 \middle| k \right) = AP\left( k \middle| k \right)A^{T} + Q\]
Correction step:
\[K(k+1) = P(k+1|k)\,\overline{C}^{T}\left( \overline{C}P(k+1|k)\overline{C}^{T} + R \right)^{-1}\]

{\small
\[
\begin{aligned}
{\widehat{x}}_{i}\left( k+1|k+1 \right) = {}& {\widehat{x}}_{i}\left( k+1|k \right) \\
 &+ K(k+1)\left( {\overline{y}}_{i}(k+1) - \overline{C}{\widehat{x}}_{i}\left( k+1|k \right) \right)
\end{aligned}
\]}
\[P\left( k+1 \mid k+1 \right) = \left( I - K(k + 1)\overline{C} \right)P(k + 1|k)\]
where \({\widehat{x}}_{i}\left( k + 1|k \right)\) represents the predicted state vector for time \(k + 1\), based on measurements up to time \(k\), and \({\widehat{x}}_{i}\left( k + 1|k + 1 \right)\) represents the corrected state vector after processing the measurements up to time \(k + 1\). The covariance matrices of the state estimation error, \(P\left( k|k \right)\) and \(P\left( k + 1|k \right)\) quantify the uncertainty of the estimate. The process noise covariance \(Q\) is not explicitly configured and is therefore assumed to be zero. Finally, the measurement noise covariance for the measured outputs is defined as:
\[R = \operatorname{diag}(n_{g},n_{g},n_{g},n_{g},n_{g})\]

\subsection{MPC-based CACC structure}\label{sec:mpc-struct}

Figure~\ref{fig:control-loop} shows a block diagram of the control loop and the interaction between the MPC and the vehicle plant. The predecessor's states are received via the wireless link with the communication delay \(\theta\). Together with the ego-vehicle's measurements, these variables are used to calculate the error dynamics \(e_{i}(t),\ {\dot{e}}_{i}(t),{\ddot{e}}_{i}(t)\) to form the measurable part of the MPC's input. To implement the acceleration-dependent spacing relaxation, the second-order filter of \eqref{eq:asr-transfer} generates the dynamic spacing adjustment \(d_{s,i}(t)\) from the ego-vehicle acceleration \(a_{i}(t)\) according to \eqref{eq:spacing-relaxation}; both \(d_{s,i}(t)\) and its rate \({\dot{d}}_{s,i}(t)\) are then directly integrated into the error dynamics \eqref{eq:state-space}.

%% file: sec/05-numerical-evaluation.tex
\section{Numerical Evaluation}\label{sec:numeval}

The following numerical analysis aims to verify the core properties of the controller architecture presented in Section~\ref{sec:controller}. Specifically, we investigate its impact on spacing and acceleration behavior at the individual vehicle level. Furthermore, we examine how these local dynamics affect the overall stability and the maximum feasible length of the platoon. Finally, a safety assessment verifies that emergency braking remains possible under all investigated conditions. A sensitivity analysis then assesses the robustness of the closed-loop response to follower-side model--plant mismatch.

\begin{figure}[!b]
  \centering
  \includegraphics[width=\columnwidth]{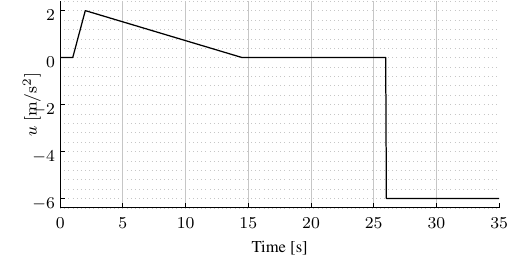}
  \caption{Demanded acceleration profile for numerical evaluation, designed to mimic the demanded acceleration of a real-world vehicle at a startup maneuver, followed by a full brake}
  \label{fig:accel-profile}
\end{figure}

For this evaluation, the lead vehicle follows the acceleration profile in Fig.~\ref{fig:accel-profile}, which mimics the dynamics of a real vehicle when the accelerator pedal is swiftly pressed to a target position and held there. Consistent with the empirical data on human startup behavior in \cite{massow_2022_numerical}, the target acceleration is set to \(a\) = \(2\,\mathrm{m/s^2}\). The profile concludes with an abrupt full-stop braking maneuver. This form allows a direct comparison with the experiments presented in Section~\ref{sec:exp}. The platoon-related parameters as defined in Section~\ref{sec:platoon-model} are summarized in Table~\ref{tab:platoon-params}. The communication frequency is chosen as \(25\,\mathrm{Hz}\), consistent with previous CACC studies \cite{oncu_2014_networkaware,vannunen_2019_stringstable,vannunen_2017_robust,ploeg_2014_thesis}.

\begin{table}[!t]
  \centering
  \caption{Platoon Related Parameter (Homogeneous Platoon). \(\phi_{i}^{d}\) and \(\theta^{d}\) are given as integer multiples of \(T_{s}\).}
  \label{tab:platoon-params}
  \footnotesize
  \begin{tabular}{@{}>{\raggedright\arraybackslash}p{\dimexpr 0.2479\columnwidth-2\tabcolsep\relax}>{\raggedright\arraybackslash}p{\dimexpr 0.1926\columnwidth-2\tabcolsep\relax}>{\raggedright\arraybackslash}p{\dimexpr 0.1919\columnwidth-2\tabcolsep\relax}>{\raggedright\arraybackslash}p{\dimexpr 0.1905\columnwidth-2\tabcolsep\relax}>{\raggedright\arraybackslash}p{\dimexpr 0.1771\columnwidth-2\tabcolsep\relax}@{}}
    \toprule
    Parameter & $\tau_{i}$ [s] & $\phi_{i}^{d}$ [--] & $\theta^{d}$ [--] & $T_{s}$ [s] \\
    \midrule
    Value & 0.1 & 4 & 1 & 0.05 \\
    \bottomrule
  \end{tabular}
\end{table}

\begin{table}[!t]
  \centering
  \caption{Controller Common Parameter}
  \label{tab:controller-common}
  \footnotesize
  \begin{tabular}{@{}>{\raggedright\arraybackslash}p{\dimexpr 0.2479\columnwidth-2\tabcolsep\relax}>{\raggedright\arraybackslash}p{\dimexpr 0.1926\columnwidth-2\tabcolsep\relax}>{\raggedright\arraybackslash}p{\dimexpr 0.1919\columnwidth-2\tabcolsep\relax}>{\raggedright\arraybackslash}p{\dimexpr 0.1905\columnwidth-2\tabcolsep\relax}>{\raggedright\arraybackslash}p{\dimexpr 0.1771\columnwidth-2\tabcolsep\relax}@{}}
    \toprule
    Parameter & $n_{g}$ & $c_f$ & $m_f$ & $k_{f}$ \\
    \midrule
    Value & 1 & 8 & 4 & 36 \\
    \bottomrule
  \end{tabular}
\end{table}

\begin{figure}[!t]
  \centering
  \includegraphics[width=\columnwidth]{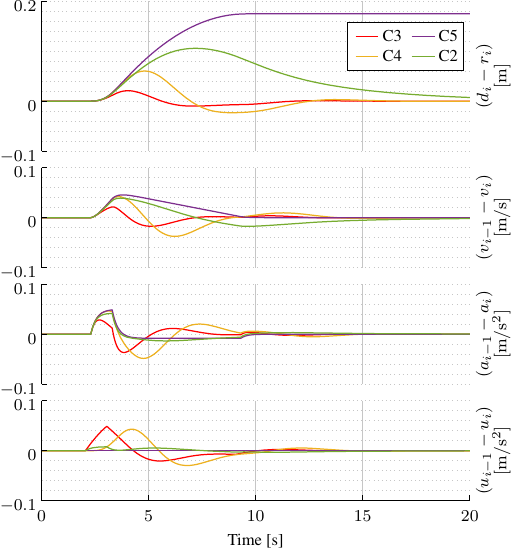}
  \caption{Numerical simulation results - comparison of controller configurations C2, C3, C4, C5 characteristics at a startup maneuver.}
  \label{fig:num-controller-comparison}
\end{figure}

\begin{table*}[!t]
  \centering
  \caption{Controller configurations used in the numerical analysis.}
  \label{tab:controller-setup}
  \scriptsize
  \setlength{\tabcolsep}{4pt}
  \begin{tabular*}{\textwidth}{
    @{\extracolsep{\fill}}
    lcccccccl
    @{}
  }
    \toprule
    Setup &
    $\delta^{d}$ &
    $K_a$ &
    $h^{d}$ &
    $w_{e1}$ &
    $w_{e2}$ &
    $w_{e3}$ &
    $w_{e4}$ &
    Main characteristic \\
    \midrule

    C2 ASR-DC & $\theta^d$ & 2 & 0 &
    1 & 2 & 2 & 1 &
    Delay compensation (DC) with acceleration-dependent spacing relaxation (ASR). \\

    C3 CDG-DC & $\theta^d$ & 0 & 0 &
    0.5 & 2 & 2 & 1 &
    Pure CDG with delay compensation (DC). \\

    C4 CDG & 0 & 0 & 0 &
    0.5 & 2 & 2 & 1 &
    Pure CDG reference without delay compensation. \\

    C5 MTG & 0 & 0 & $\theta^d$ &
    1 & 2 & 2 & 1 &
    Micro Time Gap (MTG). \\

    C6 MTG & $-\theta^d$ & 0 & $2\theta^d$ &
    1 & 2 & 2 & 0.2 &
    Increased Micro Time Gap for stronger downstream attenuation. \\

    C7 MTG & $-2\theta^d$ & 0 & $3\theta^d$ &
    1 & 2 & 2 & 0.2 &
    Further increased Micro Time Gap for stronger downstream attenuation. \\

    C8 ASR-DC & $\theta^d$ & 4 & 0 &
    1 & 2 & 2 & 1 &
    Delay compensation with stronger acceleration-dependent spacing relaxation. \\

    \bottomrule
  \end{tabular*}
\end{table*}

\subsection{Controller Comparison at the Individual Vehicle Level}\label{sec:num-single}

The investigated controller configurations are summarized in Table~\ref{tab:controller-common} and Table~\ref{tab:controller-setup}. These configurations were selected to isolate and examine individual features of the MPC approach. The reference configuration, C4, uses a CDG policy with redundant information transmission and jerk-based error dynamics. C3 extends C4 by adding delay compensation via the time shift \(\delta^{d}\), corresponding to the value of the communication delay defined for the simulation \(\delta^{d} = \theta^{d}\), i.e. one sampling step of \(0.05\,\)s. In contrast, C5 implements the Micro Time Gap with \(h^{d} = \theta^{d}\) while C6 uses \(h^{d} = {2\theta}^{d}\). These choices represent the nearest discrete increments feasible within the MPC framework, situated respectively below and above the theoretical string stability boundary derived in Section~\ref{sec:microgap-derivation}. Instead of the Micro Time Gap, C2 activates the acceleration-dependent spacing relaxation with a moderate relaxation factor of \(K_{a} = 2\), while C8 explores a stronger relaxation with \(K_{a} = 4\).

Figure~\ref{fig:num-controller-comparison} illustrates the numerical simulation results for these configurations on the first follower \(i\) following the platoon leader \(i - 1\). The plots show the physically relevant relative quantities between follower and leader, net inter-vehicle spacing \(d_{i}(t) - r_{i}\), and differences in velocity \(v_{i}(t) - v_{i - 1}(t)\), acceleration \(a_{i}(t) - a_{i - 1}(t)\), and desired acceleration \(u_{i}(t) - u_{i - 1}(t)\). As the figure shows, the jerk-based error dynamics (Section~\ref{sec:jerk}) attenuate \(u_{i}\) without the need for explicit rate weighting. This prevents the characteristic spikes in \(u_{i}\) when \(u_{i - 1}\) changes abruptly.

Compared to the reference C4, configuration C3 provides a smaller spacing deviation: compensating the communication delay (Section~\ref{sec:delay-comp}) allows an earlier response, at the price of a slightly larger transient overshoot of the desired acceleration \(u_{i}(t)\).
\begin{figure*}[!t]
  \centering
  \includegraphics[width=\textwidth]{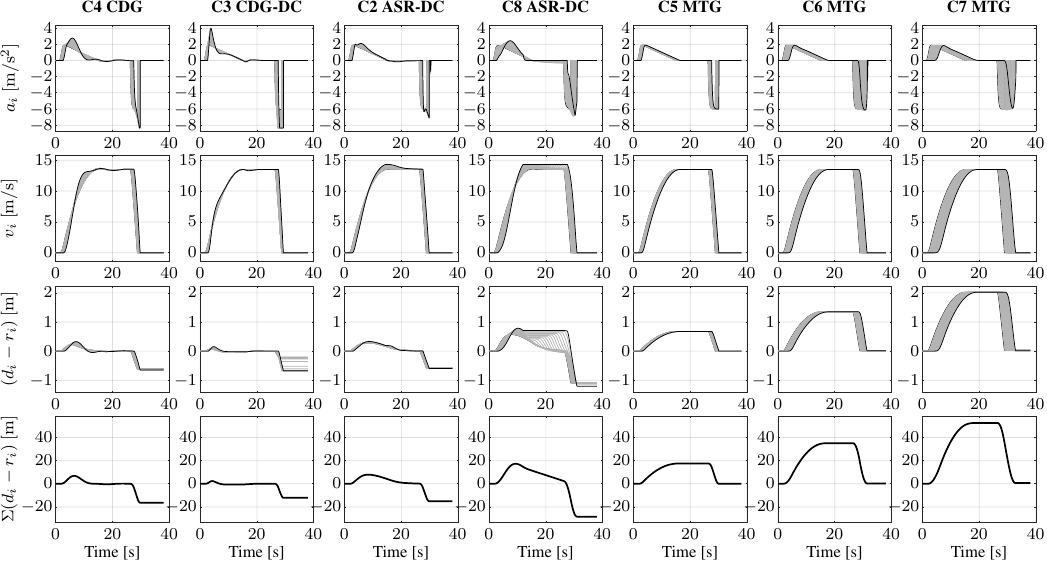}
  \caption{Numerical simulation results of all controller configuration C2-C8 within a platoon (last vehicle: black, other vehicles: grey), illustrating the net inter-vehicle spacing \(d_{i}(t) - r_{i}\), velocity \(v_{i}(t)\), and acceleration \(a_{i}(t)\).}
  \label{fig:num-platoon}
\end{figure*}

The small time shift of only \(\delta = 0.05\,\mathrm{s}\) has a measurable impact, allowing the controller to react sooner and achieve more precise distance control.

C5 (Micro Time Gap) closely reproduces the acceleration of the platoon leader, but at the cost of a larger gap at higher speeds, as expected. C2 provides an intermediate behavior: it permits a controlled, temporary increase in spacing in exchange for a smaller overshoot of the desired acceleration \(u_{i}(t)\ \). This increased spacing is then slowly compensated by a slight velocity surplus at the end of the startup phase, when the spacing converges back to the standstill distance.

\subsection{Platoon Stability and Disturbance Propagation}\label{sec:num-platoon}

In the following, we investigate how velocity and acceleration overshoots propagate through a platoon. As this work prioritizes small gaps over strict string stability, we expect these overshoots to amplify along the platoon. The goal is to identify the point where this amplification reaches physical limits or safety boundaries.

Each controller configuration (Tables~\ref{tab:controller-common} and~\ref{tab:controller-setup}) is simulated in a platoon of identical vehicles, using the parameters of Table~\ref{tab:platoon-params}. While \cite{massow_2024_closerange} suggested 20 vehicles as a maximum length, we deliberately simulate 30 vehicles here, making downstream amplification and physical limitations easier to identify and highlighting the differences between the configurations. The platoon lead vehicle follows the prescribed acceleration profile (see Fig.~\ref{fig:accel-profile}). The velocity overshoot of the simulation is limited to a maximum of \(3\,\mathrm{km/h}\) to remain within standard speed-limit tolerance thresholds (in Germany). The maximum deceleration within the simulation model is set to \(-8.33\,\mathrm{m/s^2}\), representing typical full-braking values \cite{krishnamurthy_2022_emergencybraking}, and the controller output \(u_{k}\) is constrained to the interval \([-10,4]\,\mathrm{m/s^2}\).

Although longer platoons could theoretically be realized if the lead vehicle were to start with a lower initial acceleration than \(2\,\mathrm{m/s^2}\) (see Fig.~\ref{fig:accel-profile}), we deliberately refrained from optimizing these conditions. Instead, a realistic driving profile is used as in \cite{massow_2022_numerical} to reflect the intended deployment scenario: short automated platoons starting at traffic lights, led by a human-operated vehicle that does not need to account for the vehicles behind it (OVLA communication pattern).

Figure~\ref{fig:num-platoon} shows the numerical simulation results by superimposing the temporal profiles of net inter-vehicle spacing \(d_{i}(t) - r_{i}\), velocity \(v_{i}(t)\), and acceleration \(a_{i}(t)\) for every vehicle in the platoon and the maximum of the cumulative net inter-vehicle spacing \(S_{net}^{*} = \max_{t}\ \Sigma(d_{i}(t) - r_{i})\). To quantify the downstream amplification of the desired acceleration, in a form directly related to the maximum acceleration admissible within the platoon, the Desired Acceleration Overshoot (DAO) is defined from the peak desired accelerations \(u_{i}^{\max}\) of each vehicle during startup and reported in Table~\ref{tab:dao}. The DAO is determined as \(K_{DAO}\) + 1 = \(\gamma\) with:
\[u_{i,\mathrm{max}} = \gamma\, u_{i-1,\mathrm{max}}\]

\begin{table}[!t]
  \centering
  \caption{Desired Acceleration Overshoot}
  \label{tab:dao}
  \scriptsize
  \begin{tabular}{@{}>{\raggedright\arraybackslash}p{\dimexpr 0.2260\columnwidth-2\tabcolsep\relax}>{\raggedright\arraybackslash}p{\dimexpr 0.1197\columnwidth-2\tabcolsep\relax}>{\raggedright\arraybackslash}p{\dimexpr 0.1197\columnwidth-2\tabcolsep\relax}>{\raggedright\arraybackslash}p{\dimexpr 0.1197\columnwidth-2\tabcolsep\relax}>{\raggedright\arraybackslash}p{\dimexpr 0.1341\columnwidth-2\tabcolsep\relax}>{\raggedright\arraybackslash}p{\dimexpr 0.1341\columnwidth-2\tabcolsep\relax}>{\raggedright\arraybackslash}p{\dimexpr 0.1467\columnwidth-2\tabcolsep\relax}@{}}
    \toprule
    Configuration & C2 & C3 & C4 & C8 & C5 & C6 \\
    \midrule
    \(\gamma\) & 1.007 & 1.036 & 1.018 & 1.017 & 1 & 0.999 \\
    \(K_{DAO}\) & 0.7 \% & 3.6 \% & 1.8 \% & 1.7 \% & 0 \% & -0.1 \% \\
    \(S_{net}^{*}\) [m] & 7.9 & 2.3 & 6.8 & 17.3 & 17.5 & 35.1 \\
    \bottomrule
  \end{tabular}
\end{table}

For braking, DAO is evaluated separately and analogously from the magnitudes of the minimum (most negative) control inputs. Inspired by numerical string stability analysis \cite{ploeg_2014_thesis,vannunen_2019_stringstable,oncu_2014_networkaware}, which investigates downstream error amplification in the frequency domain, we adopt the same conceptual perspective of downstream signal growth. However, as an amplification is unavoidable \cite{seiler_2004_disturbance}, DAO operationalizes the growth of the actuator-demand signal in the time domain, as a basis for sizing the feasible platoon length. Unlike error-based string-stability measures, \(K_{DAO}\) provides a time-domain engineering measure of how the actuator demand grows from one vehicle to the next and therefore links local tracking behavior to the physical feasibility of a finite mini-platoon.

\begin{itemize}
\item
  \textbf{C3 - Delay Compensation (see Section~\ref{sec:delay-comp})}: Compared to the baseline C4, C3 achieves the tightest spacing by reducing both spacing deviations and velocity oscillations. However, this configuration exhibits a consistent amplification of acceleration along the platoon, which causes it to reach the acceleration limit sooner than C4.
\item
  \textbf{C5 \& C6 - Micro Time Gap (see Section~\ref{sec:micro-gap}):} C5 yields approximately zero DAO but requires a larger inter-vehicle distance at target speed, as expected. C6 slightly attenuates the peak demand, albeit at the cost of substantially larger cumulative spacing.
\item
  \textbf{C2 \& C8 - Acceleration-Dependent Spacing Relaxation (see Section~\ref{sec:spacing-relax}):} C2 offers the best balance for maximizing platoon size at minimal spacing. The spacing relaxation reduces acceleration amplification, preventing the platoon from hitting the acceleration limit. Instead, the last vehicles in the platoon eventually reach the velocity boundary, causing their inter-vehicle gaps to grow slightly. C8 demonstrates that increasing the relaxation gain too much is counterproductive, as it fails to further damp the acceleration overshoot during startup and causes the platoon to reach the velocity limit earlier, negating the advantages over C5. This confirms that the relaxation parameter \(K_{a}\) must be selected carefully near its optimum, as demonstrated by C2.
\end{itemize}

In summary, the results demonstrate how the various controller configurations influence the inherent trade-off between minimizing the spacing error and mitigating acceleration propagation. The analysis confirms that maintaining a near-zero spacing error is often counterproductive, as it necessitates high-frequency actuation and leads to a higher amplification of acceleration peaks along the platoon. The acceleration-dependent spacing relaxation (C2) allows a larger feasible platoon size than pure CDG (C3, C4) without the high spatial requirements of classical CTG (C5, C6) at target speed. A quantitative analysis of the maximum platoon length under the ideal conditions of numerical simulation offers limited insight and is therefore deferred to the experimental validation in Section~\ref{sec:exp}.

\subsection{Safety Assessment}\label{sec:num-safety}

The simulation results in Fig.~\ref{fig:num-platoon} indicate that emergency braking remains feasible for all controller configurations at an assumed standstill distance \(r_{i}\) = \(2.95\,\mathrm{m}\), as \(d_{i}(t) - r_{i}\) does not fall below \(-1.8\,\mathrm{m}\). Despite the amplification of deceleration demand along the platoon, the minimum distance to the predecessor does not decrease further once a vehicle reaches its physical deceleration limit (\(-8.33\,\mathrm{m/s^2}\)). This behavior is a direct result of the homogeneous vehicle model assumptions and should not be interpreted as a general collision-avoidance guarantee. Note that string stability and collision-freeness are distinct properties: safety at a given gap depends on deceleration capability, delays and initial relative velocity rather than on error attenuation alone \cite{alam_2014_safeset}.

\begin{figure}[!t]
  \centering
  \includegraphics[width=\columnwidth]{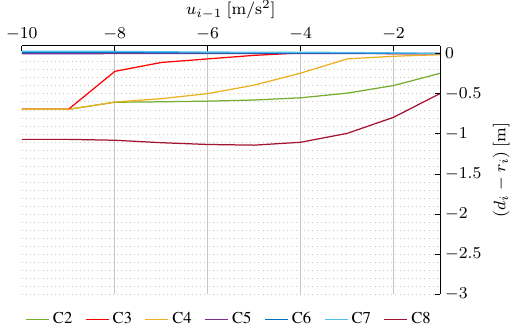}
  \caption{\textbf The resulting spacing error \(e_{i}\) after brake to standstill on the y-axis relative to the deceleration demand \(u_{i - 1}\) of the preceding vehicle on the x-axis.}
  \label{fig:spacing-error-decel}
\end{figure}

To further illustrate this, Fig.~\ref{fig:spacing-error-decel} shows a separate study in which a preceding vehicle i-1 brakes from \(50\,\mathrm{km/h}\) to a standstill using a constant step input of \(u_{i - 1}\). The graph illustrates the resulting spacing error at standstill \(e_{i}\) relative to the deceleration demand \(u_{i - 1}\) of the preceding vehicle.

C5 and C6 recover the target standstill distance in all cases because of the additional buffer introduced by the Micro Time Gap. The other configurations stop at smaller distances, with the spacing reduction saturating once the physical deceleration limit is reached. Consequently, in this idealized scenario, the braking performance of our controller does not act as a limiting factor for platoon size. In practice, the critical safety limit is defined by fleet heterogeneity, such as variances in tire grip or braking systems. C5 and C6 provide the largest safety buffer against such unmodeled real-world effects.

\subsection{Sensitivity Analysis}
\label{sec:sensitivity}

The prediction model used by the MPC is deliberately simple and cannot exactly represent the actuator dynamics of a heterogeneous legacy fleet. We therefore assess the sensitivity of the closed-loop response to follower-side
model--plant mismatch while keeping the controller, its internal prediction model, constraints, sampling time, and communication delay unchanged. The simulated follower dynamics are modified according to
\begin{equation}
\dot a_i(t)=
-\frac{1}{\tau_i}a_i(t)
+\frac{K_i}{\tau_i}u_i(t-\phi_i),
\label{eq:sensitivity-plant}
\end{equation}
where $K_i$ is the realized-to-commanded steady-state acceleration gain. Thus, $K_i=1$ corresponds to ideal realization of the commanded acceleration.

Two nested deterministic $2^3$ full-factorial designs are considered around the reference plant $(\tau_i,\phi_i,K_i)=(0.20\,\mathrm{s}, 0.20\,\mathrm{s},1)$. The inner, fleet-relevant envelope uses $\tau_i,\phi_i\in[0.15,0.25]\,\mathrm{s}$ and $K_i\in[0.90,1.10]$, whereas the outer stress envelope uses
$\tau_i,\phi_i\in[0.05,0.35]\,\mathrm{s}$ and
$K_i\in[0.85,1.15]$. Each envelope contains the eight factorial corners and shares the reference point. The controller is not re-tuned when the simulated plant parameters are changed. The analysis is performed for C2, C3, and C5, i.e., the three representative configurations subsequently emphasized in the experimental comparison. For C3, the evaluation-oriented tuning used to obtain the DAO reported in Section~\ref{sec:num-platoon} is retained. Hence, the comparison concerns the complete controller configurations, including their respective MPC tuning, rather than the spacing-policy structure in isolation.

Two complementary scenarios are considered: Scenario S1 is a conventional regulation test: both vehicles travel at $8\,\mathrm{m/s}$, the follower starts with $e_{1,i}(0)=+1\,\mathrm{m}$, and the predecessor does not accelerate. Scenario S2 instead reproduces the acceleration portion of the urban-startup profile of Fig.~\ref{fig:accel-profile}: both vehicles start from standstill, the predecessor acceleration rises to $2\,\mathrm{m/s^2}$ and subsequently returns to zero, followed by constant-speed travel. The braking phase is not
included because braking robustness is assessed separately in Section~\ref{sec:num-safety}. S1 therefore tests recovery from a spacing perturbation, whereas S2 evaluates mismatch under the maneuver directly relevant to the traffic-light application.

\begin{table*}[!t]
\centering
\caption{Sensitivity to follower-side model--plant mismatch. ``Inner'' denotes the fleet-relevant envelope and ``outer'' the wider stress envelope. S1 reports spacing-error undershoot and settling time; S2 reports the maximum spacing-error magnitude during the acceleration
phase and the maximum follower acceleration command.}
\label{tab:sensitivity-results}
\scriptsize
\setlength{\tabcolsep}{4pt}

\begin{tabular*}{\textwidth}{
@{\extracolsep{\fill}}
lcccccc
@{}
}
\toprule

& \multicolumn{3}{c}{S1: initial-error regulation}
& \multicolumn{3}{c}{S2: urban-startup acceleration} \\

\cmidrule(lr){2-4}
\cmidrule(lr){5-7}

Configuration
&
\makecell{Reference: \\
undershoot / settle}
&
\makecell{Inner: \\
worst undershoot / settle}
&
\makecell{Outer: \\
worst undershoot / settle}
&
\makecell{Reference: \\
$\max |e_1|$ / $\max |u_i|$ }
&
\makecell{Inner: \\
$\max |e_1|$ / $\max |u_i|$}
&
\makecell{Outer: \\
$\max |e_1|$ / $\max |u_i|$}
\\

\midrule

C2
&
$71.2\%$ / $8.25\,\mathrm{s}$
&
$85.0\%$ / $9.45\,\mathrm{s}$
&
$112.6\%$ / $14.65\,\mathrm{s}$
&
$0.004\,\mathrm{m}$ / $2.01\, \mathrm{[m/s^2]}$
&
$0.498\,\mathrm{m}$ / $2.22\, \mathrm{[m/s^2]}$
&
$1.091\,\mathrm{m}$ / $2.68\, \mathrm{[m/s^2]}$
\\

C3
&
$21.9\%$ / $9.25\,\mathrm{s}$
&
$26.3\%$ / $13.55\,\mathrm{s}$
&
$37.8\%$ / $14.65\,\mathrm{s}$
&
$0.064\,\mathrm{m}$ / $2.07\, \mathrm{[m/s^2]}$
&
$1.038\,\mathrm{m}$ / $2.12\, \mathrm{[m/s^2]}$
&
$2.021\,\mathrm{m}$ / $2.37\, \mathrm{[m/s^2]}$
\\

C5
&
$71.1\%$ / $8.25\,\mathrm{s}$
&
$84.7\%$ / $9.40\,\mathrm{s}$
&
$111.7\%$ / $14.60\,\mathrm{s}$
&
$0.000\,\mathrm{m}$ / $2.00\, \mathrm{[m/s^2]}$
&
$0.499\,\mathrm{m}$ / $2.21\, \mathrm{[m/s^2]}$
&
$1.087\,\mathrm{m}$ / $2.66\, \mathrm{[m/s^2]}$
\\

\bottomrule
\end{tabular*}
\end{table*}

In S1, all investigated responses recover the imposed $1\,\mathrm{m}$ spacing perturbation, with final errors below $10^{-3}\,\mathrm{m}$ over the tested parameter grid. C2 and C5 exhibit almost identical sensitivity: within the inner envelope, their worst undershoots are approximately $85\%$ and their maximum settling times remain below $9.5\,\mathrm{s}$. The C3 tuning produces
considerably less target crossing, with a worst inner-envelope undershoot of $26.3\%$, although its maximum settling time increases to $13.55\,\mathrm{s}$. Under the deliberately wider stress envelope, the responses remain bounded over the simulation horizon, but the degradation becomes more pronounced, particularly when large actuator lag and dead time occur simultaneously.

The application-oriented S2 results provide a complementary picture. At the reference point, the maximum spacing errors during acceleration are only $0.004\,\mathrm{m}$ for C2, $0.064\,\mathrm{m}$ for C3, and essentially zero for C5. Within the fleet-relevant mismatch envelope these maxima increase to $0.498$, $1.038$, and $0.499\,\mathrm{m}$, respectively. The corresponding outer-envelope values are $1.091$, $2.021$, and $1.087\,\mathrm{m}$. In all cases, the spacing error subsequently returns close to zero during the constant-speed phase; even over the outer envelope, the cruise-window mean remains within $0.020\,\mathrm{m}$ of zero. The largest commanded acceleration over all reported S2 runs is $2.68\,\mathrm{m/s^2}$, below the imposed positive
command limit of $4\,\mathrm{m/s^2}$.

The same maneuver also permits the sensitivity of DAO to be assessed. At the reference point, $K_{\mathrm{DAO}}$ is $0.74\%$ for C2, $3.66\%$ for C3, and approximately zero for C5, consistently with the relative behavior observed in
the nominal numerical evaluation. Under plant mismatch, DAO can vary substantially because the peak commanded acceleration is affected by the combined lag, delay, and acceleration-gain errors. The sensitivity results should therefore be interpreted as finite-grid evidence of practical closed-loop robustness and performance degradation, rather than as a formal robust-stability or string-stability proof. Overall, the study confirms that
the controller tolerates sizeable follower-side model mismatch, while also showing that the quantitative sensitivity depends on the selected controller parameterization and tuning.

%% file: sec/06-experimental-validation.tex
\section{Experimental Validation}\label{sec:exp}
\label{sec:exp_short}

\begin{figure*}[!t]
  \centering
  \includegraphics[width=\textwidth]{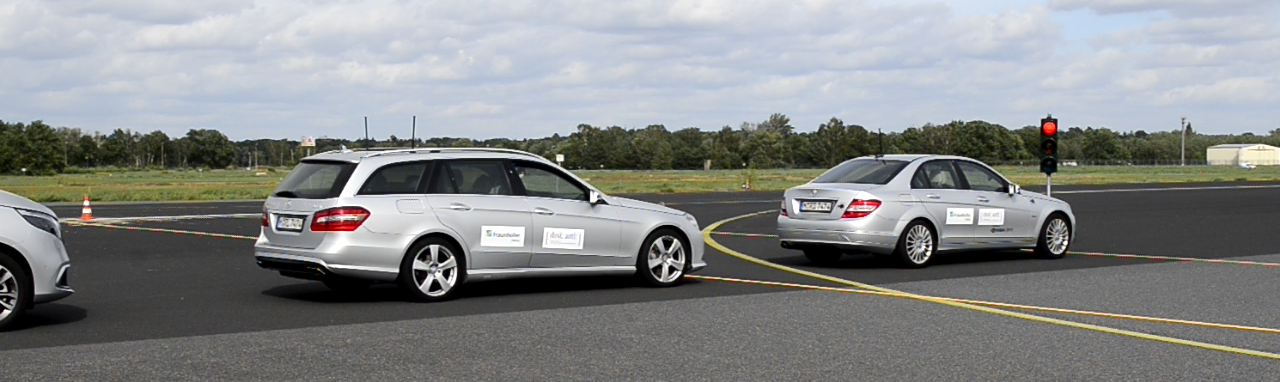}
  \caption{Experimental validation of the proposed MPC controller, retrofitted into production vehicles during real-world trials at the former Berlin-TXL airfield.}
  \label{fig:test-vehicles}
\end{figure*}

To validate the controller's performance experimentally, driving tests were conducted with production vehicles under real-world conditions on the former airfield of Berlin Tegel Airport (TXL), now part of the Urban Tech Republic (UTR), as shown in Fig.~\ref{fig:test-vehicles}. The objective of these trials is to demonstrate the controller's suitability as a retrofit solution for production vehicles and to evaluate whether the behavior identified numerically can be reproduced under external disturbances and unmodeled dynamics.

 After preliminary trials at larger distances, repeated startup runs were performed with a nominal inter-vehicle distance of \(2.95\,\mathrm{m}\). While this close proximity was initially perceived as uncomfortable by the drivers, a high level of trust in the system was established after numerous successful runs. Throughout several hundred trials, no critical situations occurred. A mechanical crash protection system installed on the bumpers for safety purposes was never triggered. For experiments involving emergency braking, a larger safety distance was maintained as a precautionary measure.

\begin{table}[!t]
  \centering
  \caption{Test Vehicle Fleet and Actuation Characteristics}
  \label{tab:vehicle-fleet}
  \scriptsize
  \setlength{\tabcolsep}{2pt}%
  \begin{tabular*}{\columnwidth}{@{\extracolsep{\fill}}lllll@{}}
    \toprule
    Vehicle & Power-train & Engine (\(\tau\) / \(\varphi\)) & Brake (\(\tau\) / \(\varphi\)) & Actuation Interface \\
    \midrule
    2013 S212 & Diesel engine & 0.2 s / 0.2 s & 0.1 s / 0.2 s & Torque request \\
    2017 W213 & EV-engine & 0.2 s / 0.2 s & 0.1 s / 0.2 s & Paravan Space Drive\textsuperscript{a} \\
    2009 W204 & Gasoline engine & 0.15 s / 0.15 s & 0.1 s / 0.15 s & Manual \\
    \bottomrule
  \end{tabular*}
  \\[2pt]
  \parbox{\columnwidth}{\raggedright\scriptsize
    \textsuperscript{a}Space Drive, PARAVAN, Germany.
    Product information:
    \url{https://www.paravan.com/product-solutions/paravan-space-drive/paravan-space-drive-system}
    (accessed 28 Aug.\ 2026).}
\end{table}

Three vehicles were used for the experiments, as detailed in Table~\ref{tab:vehicle-fleet}%
. The controller was implemented in the follower vehicle, a Mercedes-Benz S212 (E 350 CDI Diesel). A Mercedes-Benz W213 (E 350 e Hybrid) served as the lead vehicle and was operated exclusively in electric mode. Due to the hybrid powertrain, this vehicle has model parameters (\(\tau\) and \(\varphi\)) similar to those of the follower, although its acceleration characteristics during startup differ significantly due to the electric motor. A Mercedes-Benz W204 (C 250 CGI Gasoline) was used as an alternative lead vehicle to further increase variety. All three vehicles are equipped with automatic transmissions featuring torque converters but differ in gear ratios and shift points.

These diverse powertrain types result in distinct torque development behaviors during startup. This heterogeneity is intentional, as it demonstrates that precise spacing is achieved through the robustness of the developed controller rather than through vehicle similarity.

The controller was integrated into the follower vehicle using a dSPACE MicroAutoBox II \cite{vannunen_2019_stringstable} (\(20\,\mathrm{Hz}\) cycle time for the MPC and \(100\,\mathrm{Hz}\) for CAN access), replacing the original Distronic control unit (RDU). For direct communication between the vehicles, a communication unit \cite{ploeg_2011_design} was connected via the CAN interface. Aside from these components, the vehicles remained in their production state, using the standard radar system as well as the factory sensors for acceleration and wheel speed. These production signals are read over the CAN bus at different resolutions, which appears as a staircase pattern in some of the signals depicted in the following figures. The same sensing and communication set would be required for a CTG-based CACC retrofit.

\begin{figure*}[!b]
  \centering
  \includegraphics[width=0.737\textwidth]{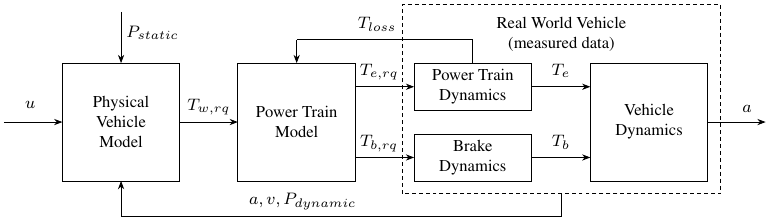}
  \caption{Simplified structure of the vehicle control model.}
  \label{fig:vehicle-control-model}
\end{figure*}

While the communication structure and controller design presented in Section~\ref{sec:platoon-model} allow for online adaptation of the predecessor's model parameters (\(\tau\) and \(\varphi\)), these parameters were kept fixed for the following experiments (Table~\ref{tab:vehicle-fleet}%
). This approach was chosen for comparability with the numerical analysis, to isolate the characteristics of the different controller configurations, and so that changes in lead-vehicle dynamics directly test the controller's robustness to systematic model mismatch.

\subsection{Vehicle Instrumentation}\label{sec:instrumentation}

The implementation of the vehicle control interface generally follows established approaches such as those found in \cite{lee_2025_openroad,flores_2023_enabling,ploeg_2011_design,nieuwenhuijze_2012_heavyduty}. However, it differs in one respect that enables a precise execution of acceleration demands.
\begin{figure}[!b]
  \centering
  \includegraphics[width=0.85\columnwidth]{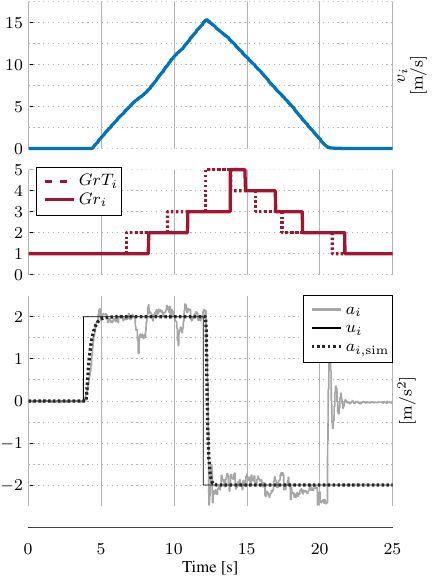}
  \caption{Validation of the actuation model - step response of acceleration requests.}
  \label{fig:actuation-validation}
\end{figure}
In contrast to many existing works that manipulate the accelerator pedal position, we directly transmit torque requests for the engine and brakes via the CAN bus (cf. \cite{massow_2018_prototyping}). This approach uses the OEM-specific torque model available in the vehicle's internal control units, which is calibrated with high precision during series development. This internal model accounts for dynamic torque losses from auxiliary units as well as internal friction and inertia effects within the powertrain. Consequently, the resulting longitudinal acceleration at the wheels can be predicted and controlled with high accuracy.

\subsubsection{Vehicle Control Model Structure}

Figure~\ref{fig:vehicle-control-model} shows a simplified structure of the vehicle control model, limited to the quantities required to follow the translation of the desired acceleration into engine and brake torque requests.

The physical vehicle model calculates the required tractive force to implement the desired acceleration \(u\) and derives the necessary wheel torque \(T_{w,rq}\). The governing equations follow the principles outlined in \cite{massow_2018_prototyping}. The model incorporates static parameters \(P_{static}\) (e.g., gear ratios, drivetrain inertia, rolling resistance, and aerodynamics) as well as dynamic variables \(P_{dynamic}\) (e.g., current velocity, selected gear, vehicle mass including fuel and payload, and estimated road gradient).

\begin{figure}[!t]
  \centering
  \includegraphics[width=0.85\columnwidth]{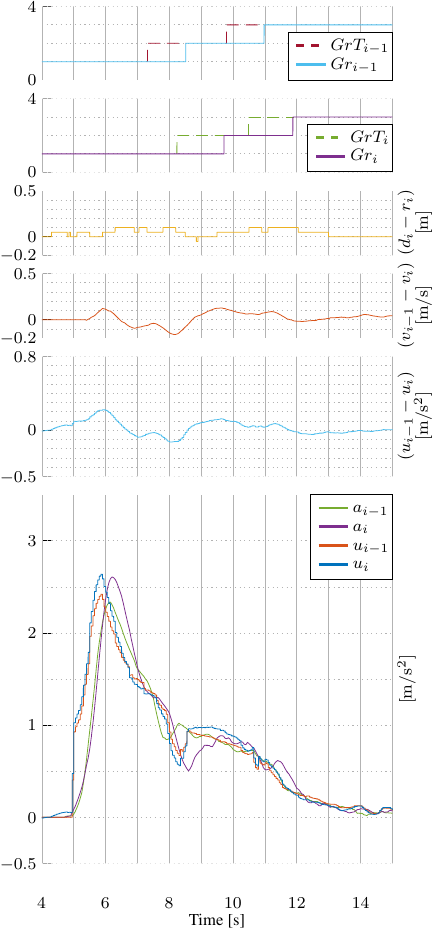}
  \caption{Acceleration profile of a single trial for setup C3.}
  \label{fig:accel-profile-c3}
\end{figure}

The calculated wheel torque \(T_{w,rq}\) is subsequently translated by the powertrain model into specific torque requests for the engine \(T_{e,rq}\) and the brakes \(T_{b,rq}\), resulting in the actual vehicle acceleration \(a\) and velocity \(v\). During this process, torque losses \(T_{loss}\) of auxiliary systems and internal drivetrain friction are explicitly considered.

\subsubsection{Vehicle Control Model Validation}

To validate this actuation model, several test runs were conducted to analyze the vehicle's response to specific acceleration profiles. Figure~\ref{fig:actuation-validation} illustrates a representative trial involving an acceleration of \(2\,\mathrm{m/s^2}\) up to \(15\,\mathrm{m/s}\), followed by a deceleration of \(-2\,\mathrm{m/s^2}\) to a standstill. Here, \({Gr}_{i}\) denotes the currently engaged gear, while the target gear \({GrT}_{i}\) represents the gear requested by the transmission control unit.

\begin{figure}[!t]
  \centering
  \includegraphics[width=0.85\columnwidth]{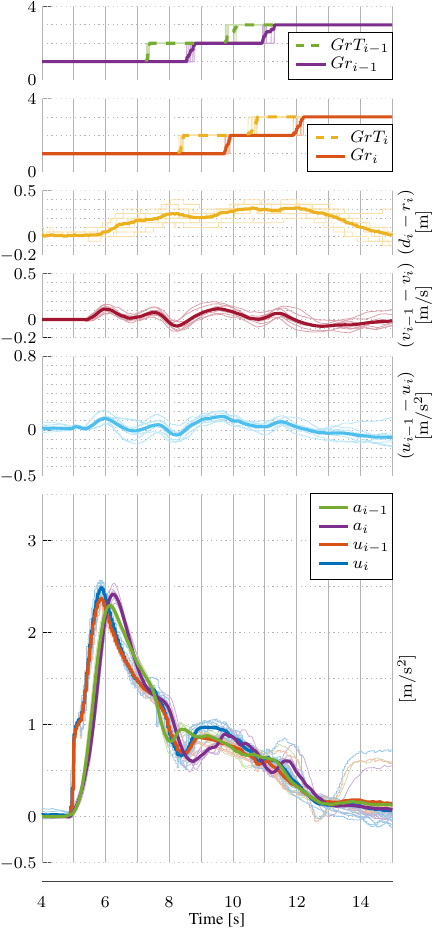}
  \caption{Averaged experimental results for the configuration C2 -- homogeneous case}
  \label{fig:homog-c2-averaged}
\end{figure}

\begin{figure}[!t]
  \centering
  \includegraphics[width=\columnwidth]{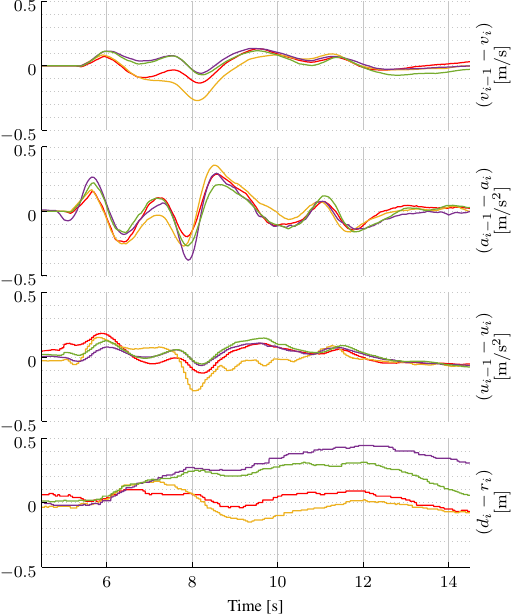}
  \caption{Comparison of all controller configurations in experimental trials confirms the characteristics from the numerical analysis -- homogeneous case (C2-green, C3-red, C4-yellow, C5-violet).}
  \label{fig:homog-comparison}
\end{figure}

 The measured vehicle response correlates well with the first-order plant model \(a_{i,sim}\) used by the MPC, comparable to results obtained with purely electric drives \cite{dehaan_2025_heterogeneous} and hybrid drives \cite{ploeg_2014_thesis}. Discrepancies occur primarily during gear shifts, which are not explicitly modeled within the MPC. These shifts lead to temporary deviations in spacing, which the MPC must compensate for as model--plant mismatches. Explicitly modeling these switching dynamics could further enhance control quality and remains a subject for future research.

\subsection{Homogeneous Platoon}\label{sec:homog}

The experimental analysis begins with a homogeneous platoon consisting of the W213 lead vehicle and the S212 follower. Here, homogeneous refers to the model parameters \(\tau\) and \(\varphi\), which are similar for both vehicles (Table~\ref{tab:vehicle-fleet}), whereas their powertrains differ. The objective is to evaluate the controller's spacing precision under real-world conditions and to verify whether the characteristics identified in the numerical analysis manifest in experimental trials when subjected to unmodeled dynamics. For a reproducible startup, the lead vehicle's accelerator pedal was stepped to 38\%, resulting in a target acceleration of at least \(2\,\mathrm{m/s^2}\), and the speed limiter of the W213 was set to \(30\,\mathrm{km/h}\).

Figure~\ref{fig:accel-profile-c3} illustrates the commanded and measured accelerations of both vehicles in addition to the relative quantities evaluated in Section~\ref{sec:num-single}, using a single representative trial from controller configuration C3. \({Gr}_{i}\ \)and \({GrT}_{i}\) refer to the current gear and the target gear of vehicle \(i\). The latter is changed by the vehicle before gear shifting is initiated. Unlike simulation data, these real-world measurements are subject to transient disturbances (e.g., sensor noise, road gradient, chassis movement, and gear shifts). For improved readability, the measured acceleration values have been smoothed. The largest discrepancies in the acceleration of both vehicles occur during gear shifts, which take place at different times for each vehicle and are not represented in the first-order plant model of the MPC. Despite these effects, the controller maintains stable spacing and exhibits a DAO of approximately 7.6\% relative to the lead vehicle.

To compare the characteristics of the different controller configurations (C2, C3, C4, C5) with the numerical analysis, the impact of stochastic disturbances was reduced by averaging multiple trials. Eight runs were performed for each configuration. All measurands were synchronized at initial movement and averaged following the evaluation methodology presented and motivated in \cite{massow_2018_prototyping}. The procedure and the resulting averaged profiles are illustrated for configuration C2 in Fig.~\ref{fig:homog-c2-averaged}. This averaging process filters out transient noise, making systematic errors visible, such as the increased spacing error just before reaching the target velocity at about \(12\,\mathrm{s}\). This operating point, marking the transition from acceleration to steady-state cruising, represents the highest uncertainty of the first-order plant model of the MPC using fixed values of \(\tau\) and \(\varphi\).

\subsubsection{Analysis of the Controller Configurations}

Figure~\ref{fig:homog-comparison} depicts a comparison of all controller configurations in experimental trials, replicating the numerical analysis in Section~\ref{sec:num-single}%
. The characteristics from the numerical analysis are confirmed: C3 maintains a nearly constant distance, while C4 exhibits larger spacing deviations. As in the simulation, C3, which compensates the communication delay while retaining pure CDG, provides the tightest spacing regulation but produces the largest DAO (7.6\% relative to the lead vehicle). By relaxing the spacing by \(0.25\,\mathrm{m}\) during startup, C2 reduces the DAO to 5.1\%, while C5 produces the smallest experimental DAO of 3.1\% (compared to 0\% in simulation). Table~\ref{tab:performance-comparison} summarizes the DAO of the averaged results as well as the variance, the minimum spacing error \(e_{1}\), and the maximum of the net inter-vehicle spacing of all test runs during the acceleration phase. All configurations exhibit a positive spacing error \(e_{1}\ \)of approximately \(0.1\,\mathrm{m}\) just before reaching the target speed in the steady-state phase, settling at approximately \(-0.1\,\mathrm{m}\) once the velocity is constant. This is caused by a systematic miscalculation of the lead vehicle's demanded acceleration during steady-state cruising. The redundant information transmission (Section~\ref{sec:redundant}) contributed to keeping this offset small.

\begin{table}[!t]
  \centering
  \caption{Performance Comparison of all Configurations}
  \label{tab:performance-comparison}
  \scriptsize
  \begin{tabular}{@{}>{\raggedright\arraybackslash}p{\dimexpr 0.3823\columnwidth-2\tabcolsep\relax}>{\raggedright\arraybackslash}p{\dimexpr 0.1544\columnwidth-2\tabcolsep\relax}>{\raggedright\arraybackslash}p{\dimexpr 0.1544\columnwidth-2\tabcolsep\relax}>{\raggedright\arraybackslash}p{\dimexpr 0.1544\columnwidth-2\tabcolsep\relax}>{\raggedright\arraybackslash}p{\dimexpr 0.1544\columnwidth-2\tabcolsep\relax}@{}}
    \toprule
    & C2 & C3 & C4 & C5 \\
    \midrule
    \(K_{DAO}\) {[}\%{]} & 5.1 & 7.6 & 6.4 & 3.1 \\
    \(\min{\left( e_{1} \right)\ \lbrack m\rbrack}\) & -0.132 & -0.100 & -0.250 & -0.125 \\
    \({var}{\left( e_{1} \right)\ \lbrack m\textsuperscript{2}\rbrack}\) & 0.0048 & 0.0050 & 0.0138 & 0.0047 \\
    \(\max\left( d_{i}(t) - r_{i} \right)\lbrack m\rbrack\) & 0.31 & 0.1 & 0.16 & 0.44 \\
    \bottomrule
  \end{tabular}
\end{table}

This steady-state offset is consistent with model--plant mismatch in the nominal MPC formulation and could be addressed by disturbance-state augmentation for offset-free tracking.

\subsubsection{Extrapolation of Platoon Size Limits}

To estimate the maximum viable platoon length, we consider the maximum acceleration threshold permitted within the platoon. We assume that the DAO is proportional to the commanded acceleration of the preceding vehicle, as observed in the numerical analysis in Section~\ref{sec:numeval}. To define a realistic upper limit for acceleration within a vehicle platoon, we reference empirical data from three studies: \cite{massow_2022_numerical} in 2012, \cite{bogdanovic_2013_acceleration} in 2006, and \cite{bogdanovic_2013_acceleration} in 2009. These studies analyzed urban startup scenarios and provide both average and upper boundary values (95th percentile) for acceleration. The findings are summarized in Table~\ref{tab:accel-limits} and indicate that short-term accelerations up to approximately \(3.3\,\mathrm{m/s^2}\) are considered acceptable. Since the peak acceleration typically occurs only within the first 1--2~s after startup (see Fig.~\ref{fig:accel-profile-c3}), we assume that this threshold is tolerable for such brief durations within the platoon.

\begin{table}[!t]
  \centering
  \caption{Realistic Accelerations Limits}
  \label{tab:accel-limits}
  \scriptsize
  \begin{tabular}{@{}>{\raggedright\arraybackslash}p{\dimexpr 0.2717\columnwidth-2\tabcolsep\relax}>{\raggedright\arraybackslash}p{\dimexpr 0.2387\columnwidth-2\tabcolsep\relax}>{\raggedright\arraybackslash}p{\dimexpr 0.2371\columnwidth-2\tabcolsep\relax}>{\raggedright\arraybackslash}p{\dimexpr 0.2524\columnwidth-2\tabcolsep\relax}@{}}
    \toprule
    Study (Year) & Average Acceleration {[}m/s\textsuperscript{2}{]} & Upper limit of P5-P95
    {[}m/s\textsuperscript{2}{]} & Number of considered situations \\
    \midrule
    \cite{massow_2022_numerical} (2012) & 2.1 & 3.1 & 3,546 \\
    \cite{bogdanovic_2013_acceleration} (2006) & 1.9 & 3.3 & 1,210 \\
    \cite{bogdanovic_2013_acceleration} (2009) & 2.1 & 3.4 & 1,433 \\
    \bottomrule
  \end{tabular}
\end{table}

Based on the experimentally determined DAO values for configurations C2 through C5, we can extrapolate the maximum permissible platoon length before the acceleration exceeds the \(3.3\,\mathrm{m/s^2}\) threshold. The maximum desired acceleration \(u_{i,\max}\) of the i-th vehicle in the platoon is calculated as:
\[u_{i,\mathrm{max}} = u_{1,\mathrm{max}}{(1+K_{DAO})}^{i - 1}\]
This method provides a simplified estimation: the tests identify local amplification using one follower, whereas propagation through longer heterogeneous platoons involves additional dynamic interactions and cumulative control delays. It nevertheless illustrates the order-of-magnitude differences in scalability between the individual controller configurations. The results are presented in Table~\ref{tab:platoon-size}. In the table, an entry of ``--'' indicates that the acceleration limit is not reached within a typical platoon size, allowing for the length of 20 vehicles suggested in \cite{massow_2024_closerange}. Recent work confirms that information flow topology and maximum platoon size jointly influence the stability of mixed traffic flow \cite{li_2025_informationflow}, supporting the treatment of the platoon size limit as a design parameter of the platooning strategy. The results show that configurations with lower DAO, particularly C2 and C5, extend the potential platoon size compared to the high-precision configuration C3. This reveals the central trade-off: excessive precision in spacing reduces the feasible platoon length. It shows the need for a controlled relaxation that is formally integrated into the spacing policy as motivated in Sections~\ref{sec:microgap-derivation} and~\ref{sec:asr-derivation}, rather than being achieved through heuristic soft-tuning of controller gains.

\begin{table}[!t]
  \centering
  \caption{Estimation of Platoon Size Limitation}
  \label{tab:platoon-size}
  \scriptsize
  \begin{tabular}{@{}>{\raggedright\arraybackslash}p{\dimexpr 0.2527\columnwidth-2\tabcolsep\relax}>{\raggedright\arraybackslash}p{\dimexpr 0.1867\columnwidth-2\tabcolsep\relax}>{\raggedright\arraybackslash}p{\dimexpr 0.1869\columnwidth-2\tabcolsep\relax}>{\raggedright\arraybackslash}p{\dimexpr 0.1869\columnwidth-2\tabcolsep\relax}>{\raggedright\arraybackslash}p{\dimexpr 0.1869\columnwidth-2\tabcolsep\relax}@{}}
    \toprule
    \multirow{3}{=}{Platoon Leader Max. Acceleration {[}m/s\textsuperscript{2}{]}}
      & \multicolumn{4}{c@{}}{Platoon Size Limitation} \\
    \cmidrule(l){2-5}
      & C2 & C3 & C4 & C5 \\
      & \(K_{DAO}\): 5.1\,\% & \(K_{DAO}\): 7.6\,\% & \(K_{DAO}\): 6.4\,\% & \(K_{DAO}\): 3.1\,\% \\
    \midrule
    2.00 & 11 & 7 & 9 & 17 \\
    1.75 & 13 & 9 & 11 & 19 \\
    1.50 & 16 & 11 & 13 & -- \\
    1.25 & -- & 14 & 16 & -- \\
    1.00 & -- & 17 & -- & -- \\
    \bottomrule
  \end{tabular}
\end{table}

\subsection{Heterogeneous Platoon}\label{sec:hetero}

\begin{figure}[!t]
  \centering
  \includegraphics[width=0.85\columnwidth]{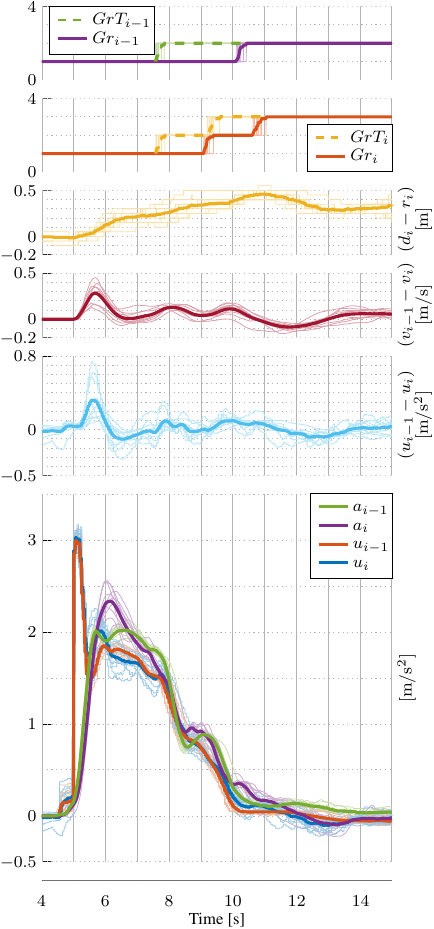}
  \caption{Averaged experimental results for the configuration C5 -- heterogeneous case.}
  \label{fig:hetero-c5-averaged}
\end{figure}

To investigate the robustness against systematic model mismatches caused by differing powertrain characteristics, the experiments were repeated using the W204 as the lead vehicle. The model parameters \(\tau\) and \(\varphi\) remained unchanged, so that the different lead vehicle produces a systematic model--plant mismatch.

\subsubsection{Dynamic Response and Actuation Differences}

Figure~\ref{fig:hetero-c5-averaged} illustrates the averaged results for configuration C5. In these trials, the lead vehicle's accelerator pedal was abruptly set to 42\%, resulting in a more dynamic startup than with the W213. The resulting demanded acceleration briefly peaks above \(3\,\mathrm{m/s^2}\) before settling at the target level of \(2\,\mathrm{m/s^2}\), leading to distinct acceleration profiles for both vehicles. While the lead vehicle requires only one gear shift, the follower shifts into third gear, causing a noticeable fluctuation in spacing just before reaching the target velocity. Despite these substantially different powertrain and actuator dynamics, the controller compensates for the mismatch with a DAO of approximately 15\%. It maintains stability and converges to the target spacing of \(d_{i} - r_{i} = 0.35\,\mathrm{m}\).

\subsubsection{Controller Comparison under Heterogeneous Conditions}

\begin{figure}[!t]
  \centering
  \includegraphics[width=\columnwidth]{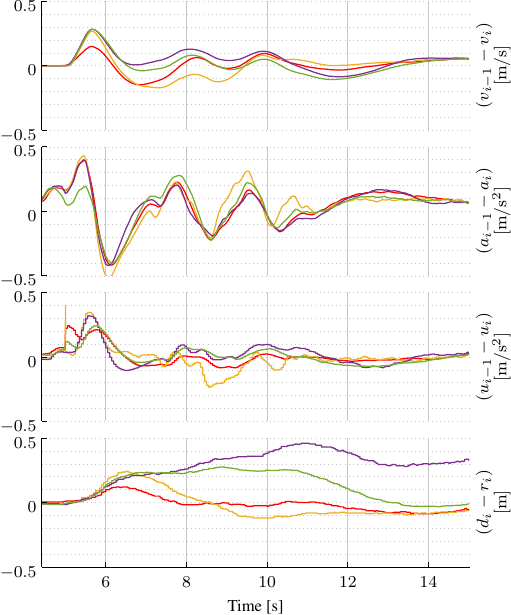}
  \caption{Comparison of all controller configurations in experimental trials -- heterogeneous case (C2-green, C3-red, C4-yellow, C5-violet.}
  \label{fig:hetero-comparison}
\end{figure}

Figure~\ref{fig:hetero-comparison} compares the different controller configurations under these conditions, as the counterpart to the homogeneous trials in Fig.~\ref{fig:homog-comparison}. All configurations initially produce a positive spacing error during startup, which is a direct consequence of the described dynamic discrepancies.

Configuration C3, with its delay compensation, responds most rapidly to these deviations. In contrast to the homogeneous case, C3 requires less DAO (10\%) than C4 (16\%), C2 (12\%), and C5 (15\%); the ranking observed in the homogeneous W213--S212 tests is thus not universal. In particular, delay compensation becomes advantageous for the faster transient of the W204 predecessor. Nevertheless, all configurations preserve bounded spacing regulation, even under these heterogeneous conditions.

\subsection{Safety Analysis}\label{sec:exp-safety}

The numerical analysis in Section~\ref{sec:num-safety} demonstrated that during emergency braking, the minimum distance to the vehicle ahead does not continue to decrease behind the first vehicle in the platoon that reaches its physical maximum deceleration. This saturation occurs regardless of the controller used or the platoon length and is a direct consequence of homogeneous model assumptions. In the real world, the lower limit for the spacing within a platoon results from the heterogeneity of the vehicle fleet, particularly differences in braking capabilities and tire--road friction.

While this heterogeneity could not be investigated experimentally due to the limited number of available test vehicles, it is possible to verify whether the controller handles dynamic inputs, such as a sudden step in demanded deceleration, in a stable manner.

\subsubsection{Experimental Safety Analysis}

\begin{figure}[!t]
  \centering
  \includegraphics[width=\columnwidth]{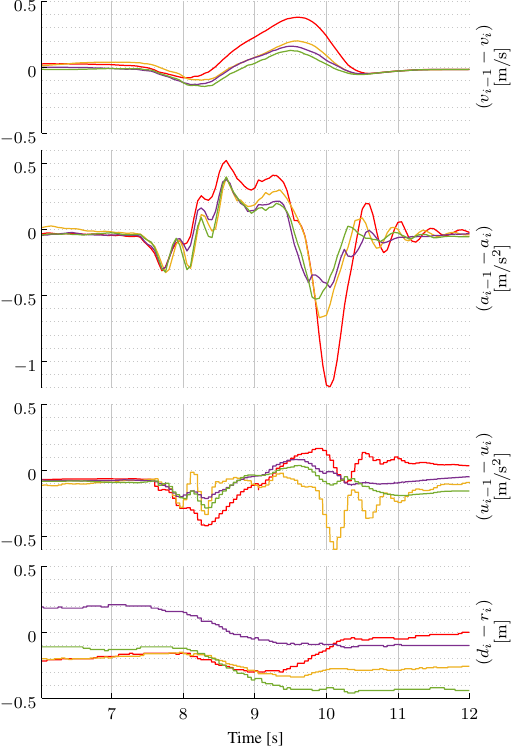}
  \caption{Comparison of braking performance across controller configurations. (C2-green, C3-red, C4-yellow, C5-violet)}
  \label{fig:braking-comparison}
\end{figure}

Braking at the physical limit was deliberately not attempted for safety reasons. Instead, the lead vehicle was subjected to a step-like braking demand of \(-4\,\mathrm{m/s^2}\) to simulate an emergency scenario. The central hypothesis is that, if the same characteristic differences between the controller configurations appear during braking as during startup and in the numerical analysis, this indicates the controller's robustness in safety-critical situations.

Figure~\ref{fig:braking-comparison} illustrates the results of this investigation, similar to the comparison of the controller configurations in a startup maneuver in Fig.~\ref{fig:hetero-comparison}. Prior to braking, all configurations exhibit a small systematic spacing error of approximately \(-0.1\,\mathrm{m}\) (see Section~\ref{sec:hetero}). During the braking maneuver, the characteristics observed in the simulation and during startup remain consistent, although disturbances prevent C5 from reaching the zero-error steady state obtained in the numerical analysis. While C5 reduces its time gap and stops with an error of approximately \(-0.1\,\mathrm{m}\), C3 compensates the initial spacing reduction and approaches zero spacing error at standstill. The effect of the delay compensation is particularly evident in comparison with C4, which stops at approximately \(-0.25\,\mathrm{m}\). C2 reaches approximately \(-0.4\,\mathrm{m}\) with a DAO of approximately 7\%. For this maneuver, C5 requires 5\% and C3 10\% DAO.

These results confirm the initial hypothesis. The observed differences in spacing errors and acceleration overshoots between configurations C2, C3, C4, and C5 follow the patterns previously identified.

\subsubsection{Safety Analysis Conclusion}\label{sec:exp-safety-conclusion}

The experimental safety analysis shows that the controller remains stable even during abrupt braking maneuvers and that the characteristic differences between the controller configurations remain visible under braking. These findings support the numerical safety analysis and confirm the controller's robustness against dynamic inputs. For real-world deployment, however, the inter-vehicle spacing must be selected to account for potential variations in braking performance, such as differences in braking systems and tire conditions.

Furthermore, a further scenario for future investigation is an emergency brake maneuver initiated during the acceleration phase (i.e., at maximum acceleration). Preliminary observations during the trials suggest that such a scenario has little effect on the final stopping distance, as the inherent positive spacing error during acceleration provides an additional safety buffer. However, as the current data set is insufficient for a comprehensive statistical evaluation, this remains a hypothesis to be tested in future work.

\subsection{Summary of Experimental Results}\label{sec:exp-summary}
The experimental evaluation validates the proposed MPC framework in production vehicles. The findings confirm that the characteristic behaviors of the controller configurations identified in the numerical analysis are reproducible in a physical vehicle platoon. The theoretical trade-offs identified in Section~\ref{sec:numeval} remain valid under physical constraints. The main results are summarized as follows:

\begin{itemize}
\item
  Despite the presence of unmodeled disturbances such as gear shifts and sensor noise, all controller configurations remained stable. Specifically, configuration C3 (delay compensation) reduced the minimum spacing error to below \(20\,\mathrm{cm}\). The controller also proved robust against systematic model mismatches within a heterogeneous platoon, consistent with the sensitivity analysis in Section~\ref{sec:sensitivity}.
\item
  Due to the aforementioned nonlinearities, the DAO values observed in the field were approximately 3--4\% higher than in the numerical analysis. While simulation results initially identified C2 as the most favorable trade-off between DAO and cumulative spacing, the experimental conditions favored C5. In the trials, C5 exhibited the lowest DAO, whereas C3 showed the highest amplification among the investigated configurations.
\item
  With the W204 as predecessor, the ranking changed: C3 required the lowest DAO and C4 the highest, i.e., delay compensation becomes advantageous for the faster transient of the W204 predecessor.
\item
  Abrupt braking maneuvers confirmed that the system handles high-deceleration steps without loss of stability and that the spacing control remains reliable.
\item
  The redundant information transmission (Section~\ref{sec:redundant}) prevented error accumulation despite model--plant mismatches in the torque calculation. While these mismatches resulted in a minor steady-state spacing error, the redundancy contributed to keeping this offset small.
\end{itemize}

In conclusion, the experimental results bridge the gap between theoretical derivation and practical application. The findings confirm that the proposed MPC framework is a robust candidate for real-world CACC deployment, provided that the nominal spacing accounts for the inherent physical heterogeneity of the vehicle fleet.

%% file: sec/07-conclusion.tex
\section{Conclusion}\label{sec:conclusion}

This work presents an MPC-based CACC controller specifically designed for retrofitting heterogeneous legacy production vehicles and demonstrates its practical feasibility in repeated real-world experiments using predecessor-only communication. The combination of a first-order prediction model with precise powertrain feedforward, based on existing OEM torque models, reduces the complexity of the MPC and avoids the need to share confidential information on vehicle characteristics among vehicles. In this way, the framework enables mini-platoons departing at traffic lights with inter-vehicle distances below 3 m and tracking errors under 20 cm. This high precision is further supported by redundant information transmission, predictive delay compensation via time shifting, and jerk-based error dynamics.

The experimental validation confirmed the controller's robustness against unmodeled disturbances and the ability to handle high-deceleration steps without loss of stability. The MPC maintained stable regulation during discrete gear shifts and across heterogeneous vehicle pairings, without requiring online parameter adaptation.

A core contribution is the comparison of two relaxation strategies: acceleration-dependent spacing relaxation (C2) and the Micro Time Gap (C5). While numerical analysis initially identified C2 as the most favorable trade-off between Desired Acceleration Overshoot (DAO) and platoon length, experimental trials at \(2\,\mathrm{m/s^2}\) showed that C2 offers only marginal benefits over C5. Extrapolated throughput analysis reveals that C2 provides a performance advantage only within narrow green-phase windows (12--15 s at 50 km/h; 15--19 s at 30 km/h). Given its larger safety buffer, C5 provides the most favorable overall compromise under the nominal experimental conditions. The heterogeneous W204--S212 experiments nevertheless show that the relative DAO ranking can change with predecessor dynamics, with C3 becoming more favorable for that particular transient. However, as this assessment relies on extrapolation, further detailed simulation remains necessary to quantify the benefits of C2 and C5.

In conclusion, this work shows that very-short-gap CDG operation for platoons departing at traffic lights does not require specialized platooning hardware. By using the designed MPC and existing vehicle interfaces, intersection capacity can be significantly increased as shown in simulation \cite{massow_2022_numerical}. This shows that intelligent control strategies can help to transform existing vehicle fleets and the present infrastructure into future automated transport.

%% file: sec/08-outlook.tex
\section{Outlook}\label{sec:outlook}

Future research will build upon the findings of this work to further enhance the robustness and traffic-level impact of the proposed system:

\begin{itemize}
\item
  \textbf{V2V Parameter Sharing:} While the presented architecture allows for the transmission of vehicle-specific parameters (\(\tau\) and \(\varphi\)), future studies should quantify the performance gains when followers directly adapt to these values from the predecessor online.
\item
  \textbf{Refined Plant Modeling:} Incorporating discrete gear-shift dynamics into the MPC state-space model could mitigate the transient spacing deviations observed in internal combustion engine vehicles during the startup phase.
\item
  \textbf{Offset-Free Tracking:} Future work should investigate offset-free tracking through disturbance-state estimation, i.e., by augmenting the prediction model with a disturbance state, to address the minor steady-state spacing offset observed in the experiments.
\item
  \textbf{V2X Robustness and Redundancy:} Developing compensation strategies for packet loss is necessary for retrofit solutions with variable antenna quality. Furthermore, augmenting V2X through Visible Light Communication or Radar-based Communication should be explored. This requires developing a consensus-based decision logic to ensure data integrity and functional safety.
\item
  \textbf{Extended Safety Validation:} Providing a comprehensive statistical analysis of emergency braking maneuvers initiated during the acceleration phase. This would test the hypothesis that the inherent spacing error during startup serves as a safety buffer across diverse fleet configurations.
\item
  \textbf{Traffic Efficiency and Intersection Capacity}: A large-scale quantitative study is needed to compare the short spacing of C2 at target speed against the potentially longer, more stable platoons enabled by C5. Future work should specifically investigate the throughput potential of C2 at lower accelerations and higher cruising speeds.

\item
  \textbf{Adaptive Information Flow Topology:} Where the surrounding traffic
  permits longer platoons and throughput at an individual intersection dominates
  --- for instance on rural arterials or at one-sidedly oversaturated
  intersections --- the predecessor-following topology could be extended to a
  predecessor--leader following (PLF) scheme, for which constant distance gap
  spacing regains string stability \cite{darbha_2020_multiple}. Switching
  between both topologies at run time \cite{he_2026_markovian} would lift the
  platoon-length limit quantified by \(K_{\mathrm{DAO}}\) where it is needed,
  while retaining the low coordination and identification effort of
  predecessor-only communication wherever it is not.
\end{itemize}

%% file: sec/90-appendix.tex
\section*{Acknowledgment}

The authors would like to thank Pietro Ferraro for proofreading the manuscript and for his helpful comments on its presentation. The authors also thank Daniel Hepp for his support in preparing and conducting the experimental test drives.